\documentclass[twocolumn,preprintnumbers,amsmath,amssymb,prb,superscriptaddress]{revtex4}

\usepackage{graphicx}   % Include figure files
\usepackage{dcolumn}    % Align table columns on decimal point
\usepackage{bm}         % bold math
\begin{document}

%\preprint{APS/123-QED}

\title{Quantum entanglement generation with surface acoustic waves}

%\altaffiliation[Also at ]{Physics Department, XYZ University.}
%Lines break automatically or can be forced with \\
%\author{Second Author}%
%\email{Second.Author@institution.edu}

\author{G. Giavaras}
\affiliation{Department of Physics, Lancaster University,
Lancaster LA14YB, England}
\affiliation{QinetiQ, St.Andrews Road,
Malvern WR143PS, England}
\author{J. H. Jefferson}
\affiliation{QinetiQ, St.Andrews Road, Malvern WR143PS, England}
\author{A. Ram$\check{s}$ak}
\affiliation{Faculty of Mathematics and Physics, University of Ljubljana and
J.Stefan Institute, 1000 Ljubljana, Slovenia}
\author{T. P. Spiller}
\affiliation{Hewlett-Packard Laboratories, Filton Road, Stoke Gifford, Bristol,
 BS34 8QZ, UK}
\author{C. J. Lambert}
\affiliation{Department of Physics, Lancaster University,
Lancaster LA14YB, England}

%\author{Charlie Author}
%\homepage{http://www.Second.institution.edu/~Charlie.Author}
%\affiliation{
%Second institution and/or address\\
%This line break forced% with \\
%}%

\date{\today}% It is always \today, today,
             %  but any date may be explicitly specified
\begin{abstract}
We propose a scheme to produce spin entangled states for two
interacting electrons. One electron is bound in a well in a
semiconductor quantum wire and the second electron is transported
along the wire, trapped in a surface acoustic wave (SAW) potential
minimum. We investigate the conditions for which the Coulomb
interaction between the two electrons induces entanglement.
Detailed numerical investigation reveals that the two electrons
can be fully spin entangled depending on the confinement
characteristics of the well and the SAW potential amplitude.
\end{abstract}

%\pacs{73.63.-b, 72.15.Qm}

%\pacs{Valid PACS appear here}% PACS, the Physics and Astronomy
                             % Classification Scheme.
%\keywords{Suggested keywords}%Use showkeys class option if keyword
                              %display desired
\maketitle

\section{INTRODUCTION}

In recent years it has become appreciated that entanglement, one
of the key fundamental features of quantum physics, lies at the
heart of numerous interesting research areas. The ability to
create entanglement between qubits in a controlled manner is a
necessary ingredient for any candidate quantum information
processing \cite{nielsen} system. Entanglement between quantum
degrees of freedom of interest and those beyond our control---the
environment---is responsible for decoherence and the degradation
of pure quantum evolution. Entanglement can exist in solids even
at thermal equilibrium \cite{ghosh,vedral} and it potentially
gives a new perspective for critical phenomena \cite{osterloh}. In
solid state systems, whether from the perspective of fundamental
quantum phenomena or their assessment as candidate quantum
processing devices, a real challenge is to establish and control
entanglement between chosen quantum degrees of freedom, whilst
avoiding decoherence due to entanglement with the relevant
environment. In this work we study, from a theoretical and
modelling perspective, the generation of entanglement between
electrons in semiconductor systems that are amenable to current
fabrication and experimental techniques.

Single electron transport (SET) in a GaAs/AlGaAs semiconductor
heterostructure using a surface acoustic wave (SAW) was
demonstrated with a very high accuracy almost a decade ago by
Shilton $et$ $al$. \cite{shilton}. Originally, the SAW-based SET
devices were investigated in the context of metrological
applications and specifically for defining a quantum standard for
the current \cite{shilton,valery_1,valery_2}. However, many other
novel applications based on this technology have been proposed
aiming to manipulate the integer number of electrons in various
ways. For example, an extension of a SAW-based SET device is a
single photon source \cite{foden} a necessary tool in quantum
cryptography \cite{phoenix,gisin}.

Barnes $et$ $al$. \cite{barnes_4,barnes_5} suggested how quantum
computations can be performed and quantum gates can be constructed
using the spins of single electrons, trapped in the SAW potential
minima, as qubits. The high SAW frequency ($\sim$ 2.7 GHz) allows
a high computation rate, which is regarded as an advantage of the
SAW-based quantum computer. The electrons are carried by the SAW
in a series of narrow parallel channels separated by tunnel
barriers. At the entrance of the channels a strong magnetic field
is applied to produce a well defined initial state for the
electrons. As the electrons are driven along the channel they can
interact with electrons in adjacent channels. The degree of
interaction may be controlled by altering the height and/or the
thickness of the barriers between the channels using surface
gates. Various readout schemes that use for example magnetic Ohmic
contacts or the Stern-Gerlach effect were proposed and described
\cite{barnes_4,barnes_5}.

This novel proposal of flying qubits has attracted a lot of
interest and theoretical work has supported its efficiency, though
quantum gates have yet to be demonstrated experimentally.
Specifically Gumbs and Abranyos \cite{gumbs} calculated the
entanglement of spins, via the exchange interaction, for two
electrons driven by SAWs in two adjacent channels. More recently
Furuta $et$ $al.$ \cite{furuta} performed detailed calculations of
the qubit dynamics when the qubits pass through magnetic fields.

In recent theoretical work Rodriquez $et$ $al.$ \cite{rodriquez}
and also Bordone $et$ $al.$ \cite{bordone_1} proposed an
experiment to observe quantum interference of a single electron
using SAWs. The proposed experiment may provide an estimation of
the electron decoherence time which is an important quantity if
these devices are to be exploited in the field of quantum
information and computation. Finally, the use of single electrons
trapped in SAW potential minima for quantum computing was
considered briefly in Ref. 11 where a more general scheme to
induce entanglement was examined in which ballistic electrons
propagate along two parallel quantum wires.

\begin{figure}
\begin{center}
\includegraphics[width=8.2cm,height=6.5cm]{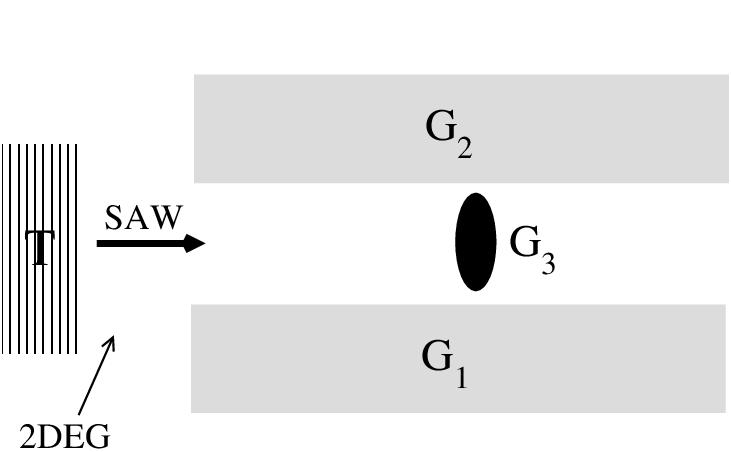}
\caption{Schematic illustration of the SAW-based device to generate spin entanglement between a static and a flying qubit. The gates $G_{1}$, $G_{2}$ define a pinched-off quasi-one dimensional channel and the gate $G_{3}$ is used to create the open quantum dot which binds the static qubit. A SAW is generated by the transducer (T) above a 2DEG and the (negative) potential on $G_{1}$ and $G_{2}$ increased until a SAW minimum contains a single electron which interacts with a bound electron under $G_{3}$.}
\end{center}
\end{figure}

In this paper, motivated by recent work on conductance anomalies \cite{rejec} and spin entanglement generation in quantum wires \cite{jefferson}, we propose a scheme to produce entangled states for two electrons utilizing SAWs. A schematic illustration of the SAW-based device is shown in Fig. 1. The SAW time-dependent potential is 
used to carry a single electron through the channel where the second electron is bound in a quantum well. The two electrons interact via the Coulomb interaction and it has been shown in various
schemes \cite{barnes_4,gumbs,jefferson,schliemann,gunlycke} that
this interaction is capable of inducing entanglement. We
investigate the conditions for which the electron in the SAW,
after passing through the region of the quantum well, will be
entangled with the electron remaining in the well. Considering the
spins of the electrons as the qubits the proposed scheme belongs
to the static-flying qubit category where specifically the qubits
interact in the same channel in contrast to \cite{barnes_4} which
involves interaction between flying qubits in different channels.

This paper is organised as follows. In Sec. II a single-electron
study is presented for the bound electron in the well and the
propagating electron in the SAW. Section III introduces the
two-electron model and considers some typical cases of
entanglement generation. In Sec. IV a Hartree approximation is
employed to explain the results of the electron dynamics. Section
V presents some general important features of the entanglement and
the main results are summarised in Sec. VI.

\section{Single electron study}

\subsection{Preliminaries}

Before examining the dynamics of the two electrons and
entanglement generation, we study the two electrons separately.
The spin of the bound electron in the well constitutes the static
qubit for the proposed scheme and therefore it is necessary to
understand how this electron behaves under the SAW propagation. In
principle, the static qubit must remain localised in the quantum
well during the computation cycle and this means that the
SAW-induced time-dependent perturbation must be such that this
condition is satisfied. It is also important that the electron in
the SAW, whose spin constitutes the flying qubit, remains bound in
the same SAW potential minimum at least up to the region where
Coulomb repulsion with the bound electron becomes important.
Although this could be achieved simply by a large SAW amplitude
the degree of screening due to the applied gate bias used to form
the quantum wire is uncertain and it may be necessary to form the
wire by an etching technique \cite{kristensen}. Finally, it is
interesting to note that well-defined single SAW pulses can be
generated \cite{barnes_4} which can be employed in order to
minimise the interaction between propagating electrons and to
allow the read-out process.

The well in the wire could be formed by surface gates, whose
geometric design and applied bias would control the confining
characteristics of the well. A single electron turnstile
\cite{kouwen} could then be used to launch an electron towards the
region of the quantum well. Whilst these aspects of realization are experimentally
feasible, details of the formation of the quantum well or the capture process are
beyond the scope of this paper.

For all the calculations in the following sections we have
employed a one-dimensional model considering only the direction of
SAW propagation, that is the positive $x$-direction. The quantum
well potential is modelled by the expression
\begin{equation}
V(x)=-V_{w}\exp\left(\frac{-x^{2}}{2l^{2}_{w}}\right),
\end{equation}
where the parameters $V_{w}$, $l_{w}$ control the depth and the
width of the well respectively. The SAW time-dependent potential
is given by \cite{robinson}
\begin{equation}
V_{SAW}(x,t)=V_{o}\{\cos[2\pi(x/\lambda-ft)]+1\},
\end{equation}
where the parameter $V_{o}$ represents the SAW potential amplitude
and to be specific we have chosen the typical values $f$=2.7 GHz
for the SAW frequency and $\lambda$=1 $\mu$m for the SAW
wavelength \cite{robinson}.

\subsection{The bound electron in the well}

In order to study the state of the electron in the well we solved
the time-dependent Schr\"odinger equation, using a Crank-Nicholson
scheme \cite{watanable} for the Hamiltonian
\begin{equation}
H_{o}=-\frac{\hbar^{2}}{2m^{*}}\frac{\partial^{2}}{\partial
x^{2}}+V_{t}(x,t),
\end{equation}
where $m^{*}=0.067m_{o}$ is the effective mass of the electron in
GaAs. The total time-dependent potential is given by the
combination of the SAW and the quantum well potential
\begin{equation}
V_{t}(x,t)=V_{SAW}(x,t)+V(x).
\end{equation}
The time evolution of the probability distribution is shown in
Fig. 2 for one SAW period $T=1/f$ and for the parameters $V_{o}=2$
meV, $V_{w}=6$ meV and $l_{w}=7.5$ nm. The quantum well parameters
have been chosen such that there is only a single bound state when
$V_{o}=0$. This becomes a quasi-bound at specific times provided
$\varepsilon_{w}<2V_{o}$, where $\varepsilon_{w}$ is the minimum
required energy to delocalize the electron from the well when
$V_{o}=0$. Although this inequality is fulfilled for the chosen
parameters, the electron still remains very well-localised in the
well for the whole SAW period as we can see from Fig. 2. This is
simply because the tunnelling time to escape from the well is much
greater than the SAW period.

\begin{figure}
\begin{center}
\includegraphics[width=7cm,height=6.5cm]{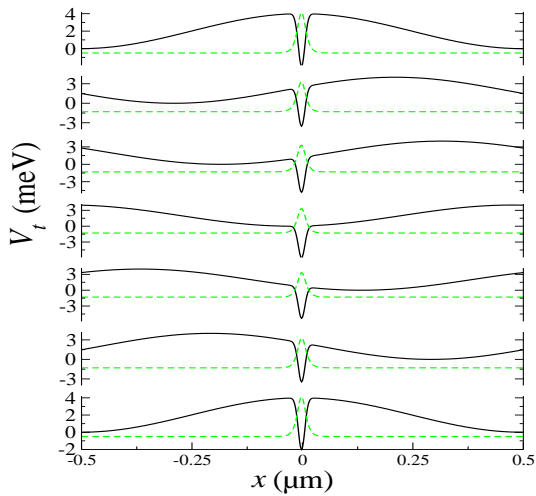}
\caption{(color online). Time evolution of the probability distribution (dashed
line in arbitrary units) of the bound state of the quantum well
and the total time-dependent potential (full line). The time
sequence is from top to bottom and specifically $t/T$=0, 0.2, 0.4,
0.5, 0.6, 0.8, 1.}
\end{center}
\end{figure}
\begin{figure}
\begin{center}
\includegraphics[width=7cm,height=6cm]{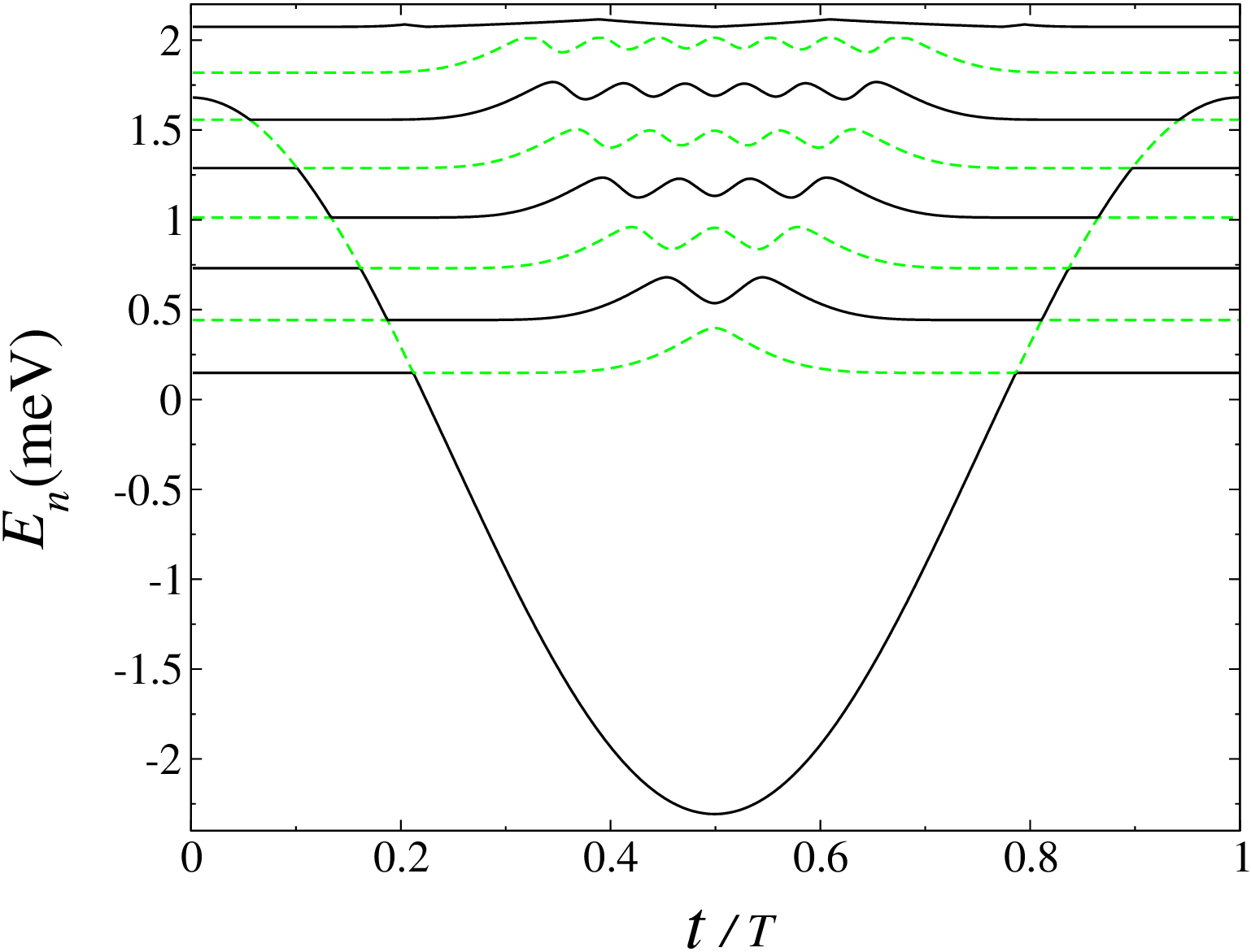}
\caption{(color online). The first few bound instantaneous eigenenergies
($E_{n}$, $n=0,1$,...) as a function of time for the total
time-dependent potential $V_{t}(x,t)$ described in the text.
Eigenenergies which correspond to an odd (even) integer are shown
with a dashed (full) line.}
\end{center}
\end{figure}

The instantaneous eigenvalues versus time, obtained by solving the
time-independent Schr\"odinger equation at each instant in time
are shown in Fig. 3 and provide insight into the dynamics. Only the first few
eigenvalues that are relevant to the time evolution process are
shown in order of increasing energy ($E_{n}$, $n=0,1$,...).
Eigenvalues corresponding to an odd (even) integer are shown with
a dashed (full) line. Note that none of the curves actually cross,
though the very small energy difference cannot be resolved in the
figure. The characteristic sine feature that develops from the
left to the right of the graph indicates that the state evolves
via non-adiabatic Landau-Zener transitions \cite{zener}. The
transition probability at an anti-crossing point, that is a point
in the graph where the two curves have minimum separation, depends
on this characteristic energy gap \cite{zener,maksym}.
Specifically, if the energy gap is large the state cannot undergo
the transition and thus it tunnels out of the well losing its
initial character, that is bound in the quantum well. On the other
hand for small energy gaps, which is the case here, the electron
can successfully undergo Landau-Zener transitions thus retaining
the initial character of its state as the time develops and the
potential profile changes. Strictly speaking, after very many SAW
cycles the electron will be delocalised from its initial quantum
well because the Landau-Zener transitions do not occur with
probability of exactly one. In our study the sine feature is only
shown for one SAW period and it describes how the energy of the
bound electron in the quantum well changes with time as the SAW
propagates. An important characteristic is that via the
Landau-Zener transitions the state retains its initial character
by changing the eigenvalue number at each anti-crossing point from
$n$ to $n\pm1$. In particular, for $t=0$ and after one SAW period
$t=T$ the SAW potential is maximum at $x=0$ and therefore the
energy level that corresponds to the bound state of the quantum
well is maximum. For $t=T/2$ the SAW potential is minimum at $x=0$
and the energy of the quantum well is the ground state energy of
the system. For $t<T/2$ the state lowers its eigenvalue number at
each anti-crossing point from $n$ to $n-1$ in order to decrease
its energy, whereas for $t>T/2$ the state increases its eigenvalue
number from $n$ to $n+1$ in order to increase its energy, via
successfully accomplishing Landau-Zener transitions. For $t>T$
this pattern of transitions is repeated. Increasing the SAW
potential amplitude and keeping the characteristics of the well
fixed the energy gaps become larger and eventually the sine
feature will disappear. In this case the electron escapes from the
quantum well tunnelling partly in the SAW potential minimum and in
the continuum. Decreasing the SAW amplitude there will be a value
such that the state of the well will be a true-bound state at all
times. In this case the state evolves adiabatically, its energy
changes sinusoidally and the electron remains localized in the
well without any effect from the SAW propagation.

To summarise for a particular quantum well there is a regime of a
small SAW potential amplitude (that satisfies
$\varepsilon_{w}<V_{o}$) where the bound electron evolves
adiabatically, followed by a regime of stronger SAW amplitude for
which the electron evolves via non-adiabatic Landau-Zener
transitions. Finally, for an even stronger SAW amplitude the
electron escapes from the quantum well. The SAW potential
amplitude must be restricted to the first two regimes for a
particular well depth. Here we consider the most interesting
intermediate case and although we only consider a quantum well
with a single bound state, it is straightforward to generalise the
results to cases with more bound states in the well.

\subsection{The propagating electron in the SAW}

In this section we study how the electron in the SAW potential
minimum propagates along the quantum wire far from the quantum
well for which we may restrict the potential of the Hamiltonian
(3) to the SAW potential only. The set of coefficients $C_{m}$,
$m=0,1...$, which satisfy
\begin{equation}
\begin{split}
\dot{C}_{m}= &-C_{m} \langle u_{m}|\dot{u}_{m} \rangle  \\
& +\sum_{n \neq m} \frac{C_{n}}{\hbar \ \omega_{mn} }\left\langle
u_{m} \left|\frac{ \partial H_{o}}{ \partial t}\right| u_{n}
\right\rangle \exp \left[ i \int^{t}_{o}
\omega_{mn}(t')dt'\right], \\
\end{split}
\end{equation}
determines the evolution of the wave function $\phi(x,t)$,
via the expansion \cite{schiff}
\begin{equation}
\phi(x,t)=\sum_{n}C_{n}(t)u_{n}(x,t) \exp
\left[-\frac{i}{\hbar}\int^{t}_{o}E_{n}(t')dt'\right],
\end{equation}
in the basis states of the instantaneous solutions
$H_{o}(x,t)u_{n}(x,t)=E_{n}(t)u_{n}(x,t)$, with $ \omega_{mn} =
(E_{m}-E_{n})/ \hbar $ the Bohr angular frequency. The wave
function $\phi(x,t)$ describes the electron in the SAW potential
minimum and satisfies the time-dependent Schr\"odinger equation.
The system of Eqs. (5) is solved using a fourth-order Runge-Kutta
method \cite{press}, although for the calculations we have dropped
the first term of Eqs. (5), since it only induces an unimportant
phase difference in the final coefficients. Figure 4 shows the
variation of the squared modulus of the coefficients, when the
initial state is the ground, the first and the second excited
state of the SAW potential minimum, ($|C^{j}_{n}|^{2}, j=n=0,1,2 $
where the superscript $j$ indicates the corresponding initial
state) as a function of time for two SAW periods and for a SAW
amplitude of $V_{o}=4$ meV. As we can see, the electron remains to
a very good approximation in the initially populated state of the
SAW minimum throughout the time evolution and furthermore the
corresponding moduli of the expansion coefficients present an
oscillating behavior.

\begin{figure}
\begin{center}
\includegraphics[width=8cm,height=6.5cm]{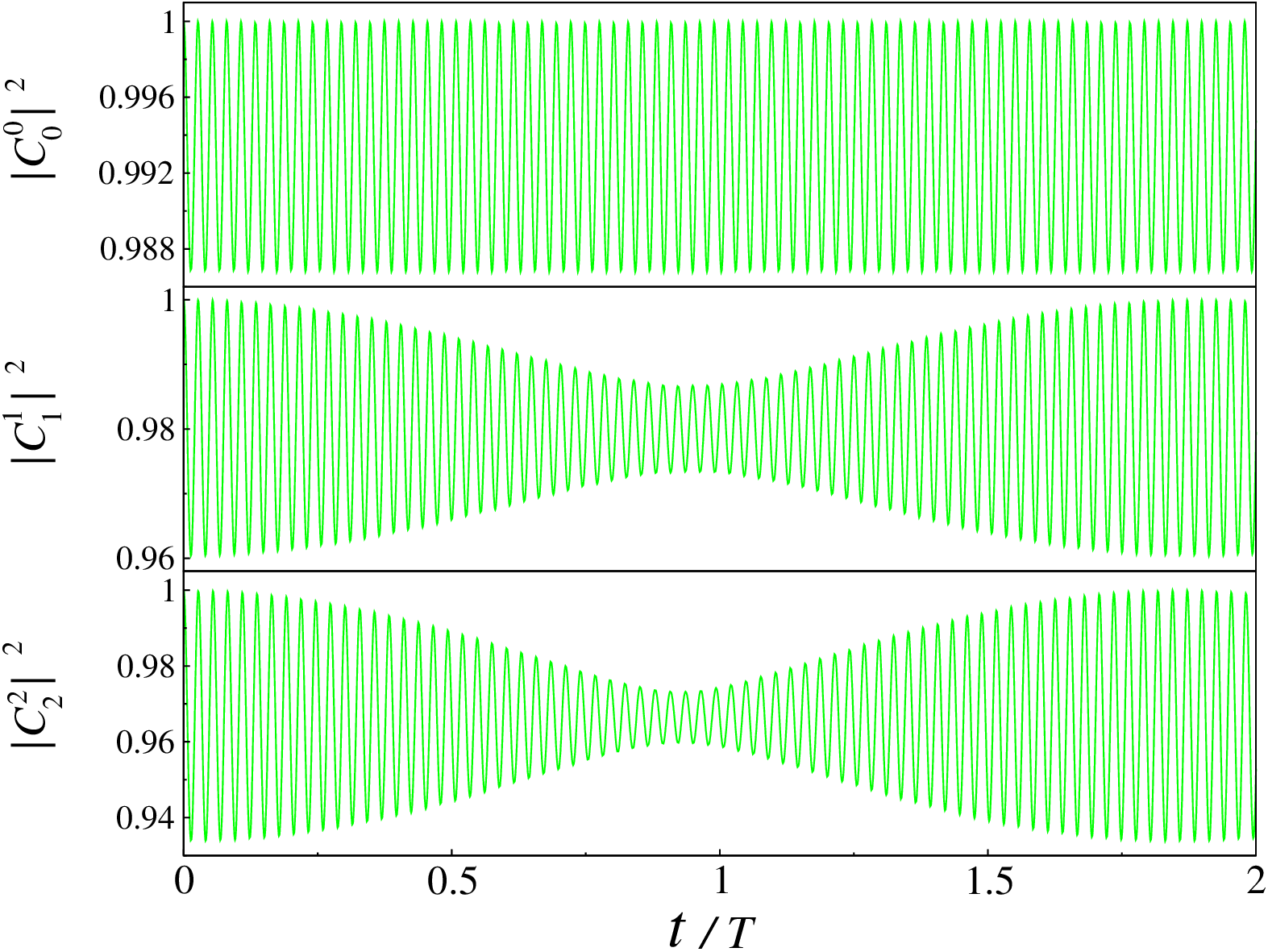}
\caption{(color online). From top to bottom the initial electron state (for
$t=0$) is the ground, the first, and the second excited state of a
SAW potential minimum respectively. The plots show how the
corresponding probability for each initial state evolves with time
for two SAW periods.}
\end{center}
\end{figure}

This behavior may be explained within the adiabatic approximation
\cite{schiff}. Starting with $C^{j}_{n}(t=0)= \delta_{nj}$,
(initial state $u_{j}$) and assuming that all the coefficients in
(5) remain constant with time $C^{j}_{n}(t>0) \approx
\delta_{nj}$, we obtain the approximate formula for all $ m \neq j$
\begin{equation}
\dot{C}^{j}_{m} \approx \frac{\omega\Lambda_{mj}}{\hbar \
\omega_{mj} }\ \exp \left[ i \int^{t}_{o}
\omega_{mj}(t')dt'\right],
\end{equation}
where we have set $\Lambda_{mj}=\langle u_{m} |\frac{\partial
H_{o}}{ \partial t}| u_{j} \rangle/\omega$, with $\omega=2\pi f$
the SAW cyclic frequency. The matrix elements $\Lambda_{mj}$ and
the frequencies $\omega_{mj}$ are time-independent and hence the
final expression for the squared modulus of the coefficients for $
m \neq j $ becomes
\begin{equation}
|C^{j}_{m}(t)|^{2} \approx \frac{ \omega^{2}|\Lambda_{mj}|^{2}}{\hbar^{2} \
 \omega^{4}_{mj}} \
4\sin^{2}\left(\frac{\omega_{mj}t}{2}\right).
\end{equation}
For the SAW potential given by (2) the matrix elements are real
for bound states and therefore $\Lambda_{mj}=\Lambda_{jm}$. Also,
the transitions are allowed when $\Lambda_{mj}\neq0$ which occurs
when $m+j$ is odd. If $\omega |\Lambda_{mj}|
\ll\omega_{mj}^{2}\hbar $ then $|C^{j}_{m}| \sim 0 $ and the
electron remains at all times in the initial state $u_{j}$. This
limit corresponds to the adiabatic approximation and is satisfied
when the system changes very slowly compared to the transition
frequency $\omega_{mj}$ associated with the states. In general,
the higher the states the smaller the transition frequency between
them and as a result the less valid the adiabatic approximation.
This can be seen directly from Fig. 4 by observing the minimum
magnitude of the oscillations which gives the maximum deviation
from the initial state and hence the deviation from the adiabatic
approximation. On the other hand, the lower the SAW frequency the
better the adiabatic approximation. In the extreme limit of a
'frozen' wave, $\omega=0$, the states become stationary acquiring
only a phase.

For the ground state, in the time interval of interest, the
adiabatic approximation is excellent. During the time evolution
there is a very small contribution from excited states and
$|C^{0}_{0}|^{2} \approx 1-|C^{0}_{1}|^{2}$ with $|C^{0}_{1}|^{2}$
given by Eq. (8) with $m=1$ and $j=0$. Hence, the frequency of the
oscillations is equal to $\omega_{10}$ and the amplitude depends
on the quantity $\omega^{2}|{\Lambda}_{10}|^{2} / (\hbar^{2} \
\omega^{4}_{10}) $. In the evolution of the first excited state
there is some contribution not only from the ground but also from
the second excited state. We may approximate the numerical results
by $ |C^{1}_{1}|^{2} \approx 1-|C^{1}_{0}|^{2}-|C^{1}_{2}|^{2}$
and using Eq. (8)
\begin{equation}
|C^{1}_{1}(t)|^{2} \approx 1-D_{01}\sin^{2}\left(\frac{\omega_{01}
t}{2}\right)- D_{21}\sin^{2}\left(\frac{\omega_{21} t}{2}\right),
\end{equation}
where the auxiliary constant is
$D_{m,j}=4\omega^{2}|{\Lambda}_{mj}|^{2}/ (\hbar^{2}
\omega^{4}_{mj})$. Defining the quantities
$\delta=(\omega_{10}-\omega_{21})/2$,$\delta_{o}=(\omega_{10}+\omega_{21})/2$
and $D=(D_{21}-D_{10})/2$, $D_{o}=(D_{21}+D_{10})/2$ we may
further write
\begin{equation}
\begin{split}
|C^{1}_{1}(t)|^{2} \approx & 1-D_{o}+D_{o}\cos(\delta_{o}t)\cos(\delta t) \\
                                  & +D\sin(\delta_{o}t)\sin(\delta t), \\
\end{split}
\end{equation}
which explains the sinusoidal variation of the amplitude in the
oscillations with a period equal to $2 \pi /\delta $. This
peculiar form is due to the very small difference between the Bohr
frequencies which define the frequency of the oscillations and the
small difference between the matrix elements which control the
amplitude of the oscillations. A similar situation occurs for the
evolution of the second excited state as shown in the bottom frame
of Fig. 4. All the relevant quantities can be defined similarly,
considering that the small deviation from this state is mainly due
to transitions to the first and the third excited state.

To summarise, the electron in the SAW propagates adiabatically
along the quantum wire even for a relatively low SAW amplitude. In
other words, it remains well-localised in the particular SAW
potential minimum as it is driven towards the bound electron in
the well, which is the main requirement for the entanglement
generation described in the next section.

\section{Electron dynamics and entanglement}

\subsection{The two-electron model}

The dynamics of the two-electron system is governed by the
time-dependent Schr\"odinger equation, with the Hamiltonian
\begin{equation}
H=\sum_{i=1,2}\left[
-\frac{\hbar^{2}}{2m^{*}}\frac{\partial^{2}}{\partial
x^{2}_{i}}+V_{t}(x_{i},t)\right]+V_{c}(x_{1},x_{2}).
\end{equation}
The single electron term $V_{t}(x,t)$ was described in Section IIB
and the Coulomb term $V_{c}(x_{1},x_{2})$ is modelled by the
quasi-one-dimensional form
\begin{equation}
V_{c}(x_{1},x_{2})=\frac{q^{2}}{4\pi \epsilon
_{r}\epsilon _{o}\sqrt{(x_{1}-x_{2})^{2}+\gamma_{c}^{2}}},
\end{equation}
where $\epsilon _{r}=13$ is the relative permittivity of GaAs.
This simplified form of the Coulomb interaction assumes that all
excitations take place in the $x$-direction, whereas in the other
two directions the electrons occupy at all times the corresponding
ground states (transverse modes). This is a good approximation
provided that the parameter $\gamma_{c}$ that models the
confinement lengths in the $y$ and $z$ directions is relatively
smaller than the confinement length scales in the $x$-direction.
For all the calculations we choose $\gamma_{c}=20$ nm, for which
the restriction to lowest transverse modes is an excellent
approximation.

The resulting two-electron time-dependent Schr\"odinger equation
is solved numerically with an explicit scheme based on a finite
difference method, which is described in detail for the case of a
single electron by Visscher \cite{Visscher}. The extension for two
electrons is straightforward.

For the initial state at $t=0$ we choose one electron to have spin
up in the ground state of the SAW potential minimum, $\psi(x)$,
and the other electron to have spin down in the ground state of
the quantum well, $\varphi(x)$. It is important to note that
$\psi(x)$ and $\varphi(x)$ are exactly orthogonal, with no spatial
region of overlap, i.e.  $\psi$ peaks around the region where the
particular SAW potential minimum is located (far from the quantum
well) and $\varphi$ peaks around $x\sim0$ where the quantum well
is located. To study the dynamics of the electrons, at time $t>0$
it is necessary to take into account the fact that the electrons
are fermions and thus indistinguishable. This is important when
the electron carried by the SAW interacts with the electron in the
quantum well, giving rise to a spin exchange interaction. The
initial state is thus represented by the Slater determinant
\begin{equation}
\Psi_{\uparrow
\\\downarrow}(x_{1},x_{2},0)=\frac{1}{\sqrt{2}}\left|
\begin{array}{cc}
      \psi(x_{1}) \chi_{\uparrow}(1)      &  \varphi(x_{1}) \chi_{\downarrow}(1)\\
      \psi(x_{2}) \chi_{\uparrow}(2)      &  \varphi(x_{2}) \chi_{\downarrow}(2)\\
\end{array}
\right|.
\end{equation}
This state is unentangled according to the criteria of Ref. 31.
Note that $\Psi_{\uparrow\downarrow}(x_{1},x_{2},0)$ can also be
expressed as a combination of a singlet and a $S_{z}$=0 triplet
state
\begin{equation}
\Psi_{\uparrow
\downarrow}(x_{1},x_{2},t)=\frac{1}{\sqrt{2}}[\Psi^{S}_{\uparrow
\downarrow}(x_{1},x_{2},t)+\Psi^{T}_{\uparrow\downarrow}(x_{1},x_{2},t)],
\end{equation}
which is also the general form of the total two-electron wave
function at all times, due to the fact that the Hamiltonian Eq.
(11) contains no spin-dependent terms. Furthermore, for the case
of two electrons, the orbital and spin parts factorize, i.e.
$\Psi^{S}_{\uparrow\downarrow}(x_{1},x_{2},t)=\Phi^{S}(x_{1},x_{2},t)\chi^{S}_{\uparrow\downarrow}(1,2)$
and similarly
$\Psi^{T}_{\uparrow\downarrow}(x_{1},x_{2},t)=\Phi^{T}(x_{1},x_{2},t)\chi^{T}_{\uparrow\downarrow}(1,2)$.
With this notation the spin components are given by
$\chi^{S/T}_{\uparrow\downarrow}(1,2)=[\chi_{\uparrow}(1)\chi_{\downarrow}(2)\mp\chi_{\downarrow}(1)\chi_{\uparrow}(2)]/\sqrt{2}$
(with the negative sign for the singlet) and the corresponding
orbital components at $t=0$, by
$\Phi^{S/T}(x_{1},x_{2},0)=[\psi(x_{1})\varphi(x_{2})\pm\varphi(x_{1})\psi(x_{2})]/\sqrt{2}$
(with the positive sign for the singlet). For $t>0$, the spin
eigenstates are unchanged whereas the orbital states are given
directly by the solution of the time-dependent Schr\"odinger
equation. The form of the orbital components at $t=0$ implies that
the two electrons do not interact, i.e. they are well separated
with negligible Coulomb interaction, and therefore are written as
symmetric and antisymmetric products of non-interacting
single-electron states. Finally, in this work we consider only
cases where the electron in the quantum well is well-localized
before ($t=0$) and after the scattering event ($t=t_{f}$), when
the electrons are well separated. The energy parameters (SAW
amplitude and quantum well characteristics) are thus chosen such
that the final electron probability distribution in the well is to
a good approximation the same as before interaction. However, this
restriction is not imposed on the propagating electron in the SAW,
which can gain energy due to a combination of the effect of the
time-dependence of the SAW potential and Coulomb repulsion.

\subsection{Entanglement measure}

In this paper, concurrence will be used as a measure of spin
entanglement. An expression for concurrence may be obtained using
the form suggested by Wooters\cite{wooter}, starting with the
total density matrix for the pure scattering state and integrating
over the orbital degrees of freedom. For the axially symmetric problems considered here, for which total spin projection along
the quantisation axis is conserved, concurrence is physically related to spin-spin correlation functions for the two 
domains $A$ and $B$ and takes the form\cite{concur} $C=2|\langle S^{+}_{A}S^{-}_{B}\rangle|$, 
where $S^{\pm}$ are the usual spin flip operators. Equivalently in terms of the symmetric (singlet) and
antisymmetric (triplet) orbital states concurrence is given by the formula\cite{concur}
\begin{equation}
C(t)=\frac{1}{N(t)}|\int_{A,B}dx_{1}dx_{2}\Phi^{\ast}_{-}
(x_{1},x_{2},t)\Phi_{+}(x_{1},x_{2},t)|
\end{equation}
where $\Phi_{\pm }(x_{1},x_{2},t)=\Phi^{S}(x_{1},x_{2},t)\pm
\Phi^{T}(x_{1},x_{2},t)$ and the normalization constant $N$ equals
\begin{equation}
N(t)=\int_{A,B}dx_{1}dx_{2}(|\Phi^{S}(x_{1},x_{2},t)|^{2}+|\Phi^{T}(x_
{1},x_{2},t)|^{2}).
\end{equation}
In these expressions the regions of integration $A$, $B$  are
chosen to be regions which the electrons are expected to occupy
before and after scattering with $A$ being the domain of one
electron and $B$ the domain of the other. Physically, the regions
$A$ and $B$ can be viewed as (position) measurement domains. For
example, sensing of the presence of an electron charge\cite{delft1,delft2,delft3} with
sufficient positional information only to identify it as being
located in some region could correspond to such a ``fuzzy''
position measurement. Since the quantum well always contains at
least one electron, we choose $A$ to be  this region, i.e.
$A$=[$x_{l}$, $x_{r}$], where $x_{l}$ and $x_{r}$ denote the
points where the bound state of the electron in the quantum well
has decayed to zero at the left and right respectively. For the
numerical calculations, these points were chosen to correspond to
a value of approximately 10$^{-4}$ of the probability density at
the peak. The region $B$ is chosen to correspond to the region of
occupation of the propagating electron. For the two-electron
scattering problem under study, we may choose this to be the total
domain excluding the well, i.e. $B$=[-$L$, $x_{l}$]$\cup$[$x_{r}$,
$L$], where [-$L$, $L$] defines the total region of space within
which the electron dynamics is studied. Note that the
corresponding concurrence is really only meaningful when the
electrons are well separated, before and after scattering, i.e. at
sufficiently small or large $t$, though it may be calculated at
any time. We refer to this concurrence as the total concurrence,
$C(t)$, .  We may also define (the potentially more useful)
reflected or transmitted concurrence, for which the measurement
domains are restricted to either the left or the right of the
quantum well respectively, i.e.  $B$=[-$L$, $x_{l}$], or
$B$=[$x_{r}$, $L$], with corresponding concurrences $C_{r}$ and
$C_{t}$. A ``fuzzy'' position measurement (charge sensing) which
gives sufficient information to resolve the outgoing electron as
reflected or transmitted could project the two electrons into a
state with the associated concurrence $C_{r}$ or $C_{t}$.  It is
also useful to define the quantities $P^{S/T}_{t}$ and
$P^{S/T}_{r}$ as the transmitted and reflected probabilities for
singlet and triplet states. The maximum probability in the whole
space of either state equals 0.5 due to the general form (14) of
the wave function. The transmitted and reflected concurrence are
considered only when the corresponding singlet or triplet
probabilities are not negligible. For the numerical calculations
the minimum limit was taken to be approximately 10$^{-2}$. We
should also mention that by definition \cite{concur} $0\leq
C\leq1$, where the limit $C=0$ corresponds to an unentangled state
and $C=1$ to a fully entangled state. For the time-dependent
problem under study, the concurrence is also time-dependent and it
is easily verified that for the initial state, $C(t=0)=0$.

\subsection{Entanglement generation}

\begin{figure}
\begin{center}
\includegraphics[width=7cm,height=6cm]{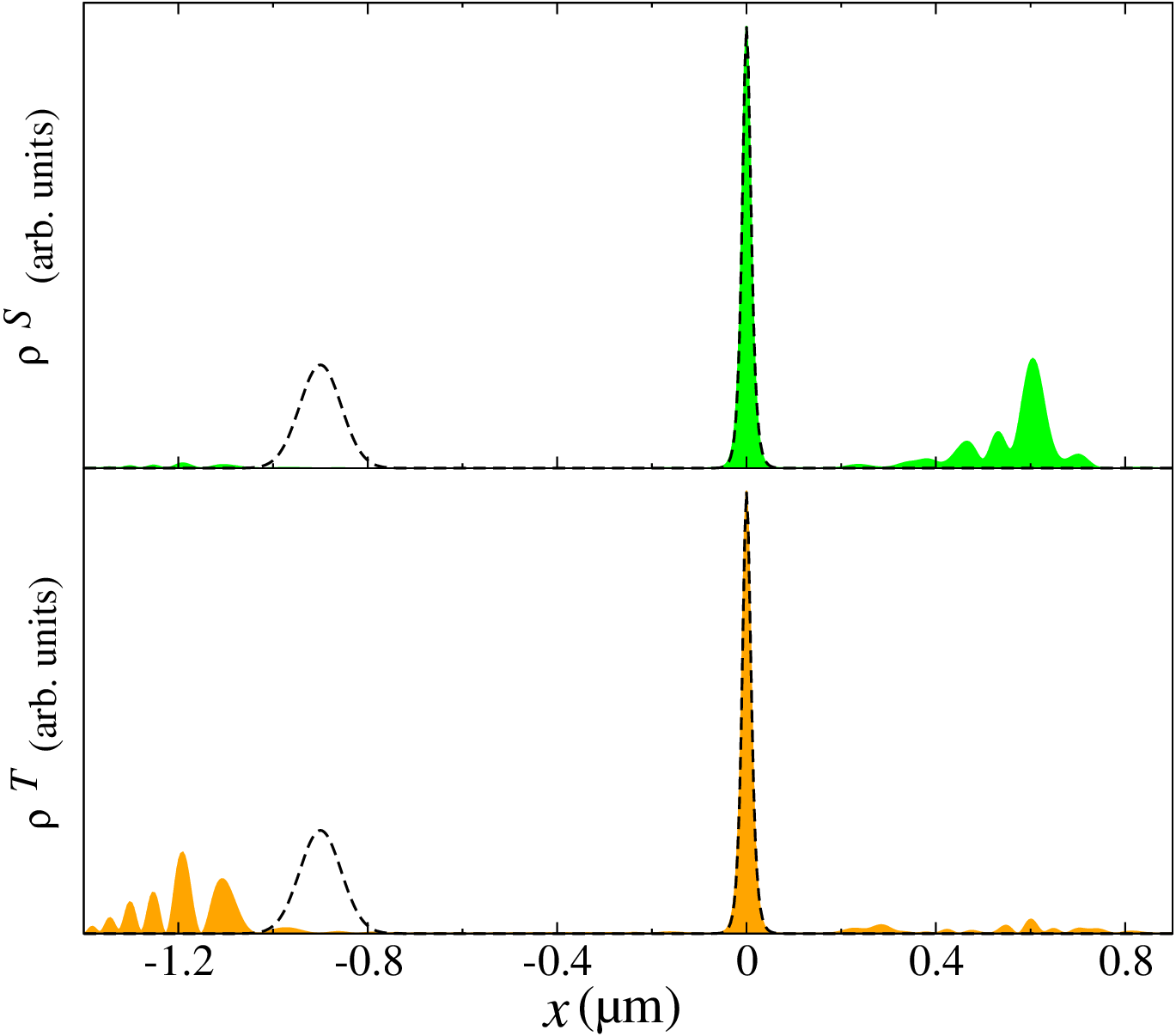}
\caption{(color online). Initial (dashed) and final electron distribution (full)
in arbitrary units when the singlet state (top) is mostly
transmitted and the triplet state (bottom) is mostly reflected.}
\end{center}
\end{figure}
\begin{figure}
\begin{center}
\includegraphics[width=7cm,height=9.5cm]{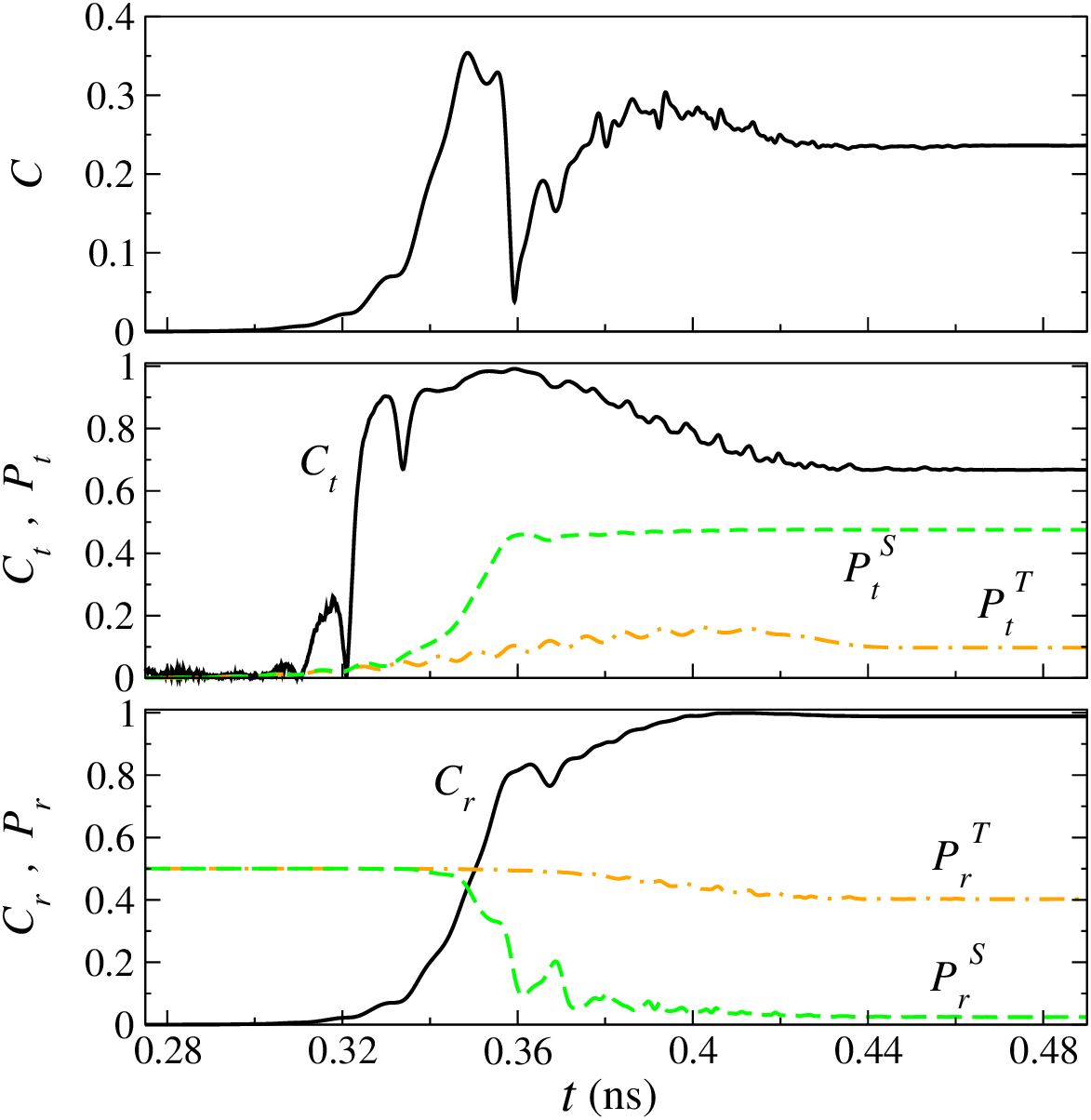}
\caption{(color online). Concurrence and relative probabilities as a function of
time when the singlet state is mostly transmitted and the triplet
state is mostly reflected.}
\end{center}
\end{figure}

In this section we present some typical scattering results that
take place when the two electrons interact via the Coulomb
interaction and demonstrate how the entanglement develops with
time due to this interaction.

Figure 5 shows the initial and final electron density
$\rho^{S/T}(x,t)=2\int dx^{'}|\Phi^{S/T}(x,x^{'},t)|^{2}$, and
Fig. 6 shows how the concurrence and the relative probabilities
develop with time for the parameters $V_{w}=6$ meV, $l_{w}=7.5$ nm
and $V_{o}=2$ meV. For these parameters the quantum well can
accommodate only a single bound electron, a second electron being
delocalised due to the Coulomb interaction. From the figure we see
that there is a very high transmission for the singlet state and
high reflection for the triplet state after scattering. The
concurrence ($C, C_{t}, C_{r}$) varies with time, due to the
interactions of the wave packets mediated by the Coulomb
interaction, and eventually saturate to a constant value when the
overlap is once again negligible. The transient time interval is
not of main interest since the degree of entanglement is important
after the scattering process when the two electrons are well
separated. Note that the reflected concurrence is close to unity
since at the left hand side the reflection probability of the
singlet state is very small compared with that of the triplet, and
a pure $S_{z}=0$ triplet is fully spin entangled. In this case,
the reflection process may be regarded as a filtering process in
which the singlet part of the initial wave function is essentially
removed by transmission to the right. One the other hand, if we
look in transmission, although the singlet part is almost fully
transmitted, the transmission of the triplet is not negligible and
interference results in an asymptotic concurrence which is
somewhat less than unity. We also see that the maximum concurrence
is also significantly reduced when the measurement domain includes
both transmitted and reflected parts after scattering.

In Fig. 7 we present results for a smaller SAW amplitude
($V_{o}=0.5$ meV) for which the initial and final electron density
of both singlet and triplet states are almost totally reflected,
the electron carried by the SAW being almost completely reflected
by the Coulomb repulsion with the bound electron. Fig. 8
illustrates how the concurrence develops in time, the concurrence
of the reflected part and that over the whole domain being
approximately the same due to the high reflection. We can see
again that the concurrence builds up with time due to the Coulomb
interaction and saturates to a constant value after reflection.
Note however, that this asymptotic value is much smaller, due
essentially to the Coulomb repulsion inhibiting significant
overlap of the wave packets.

\begin{figure}
\begin{center}
\includegraphics[width=7cm,height=6cm]{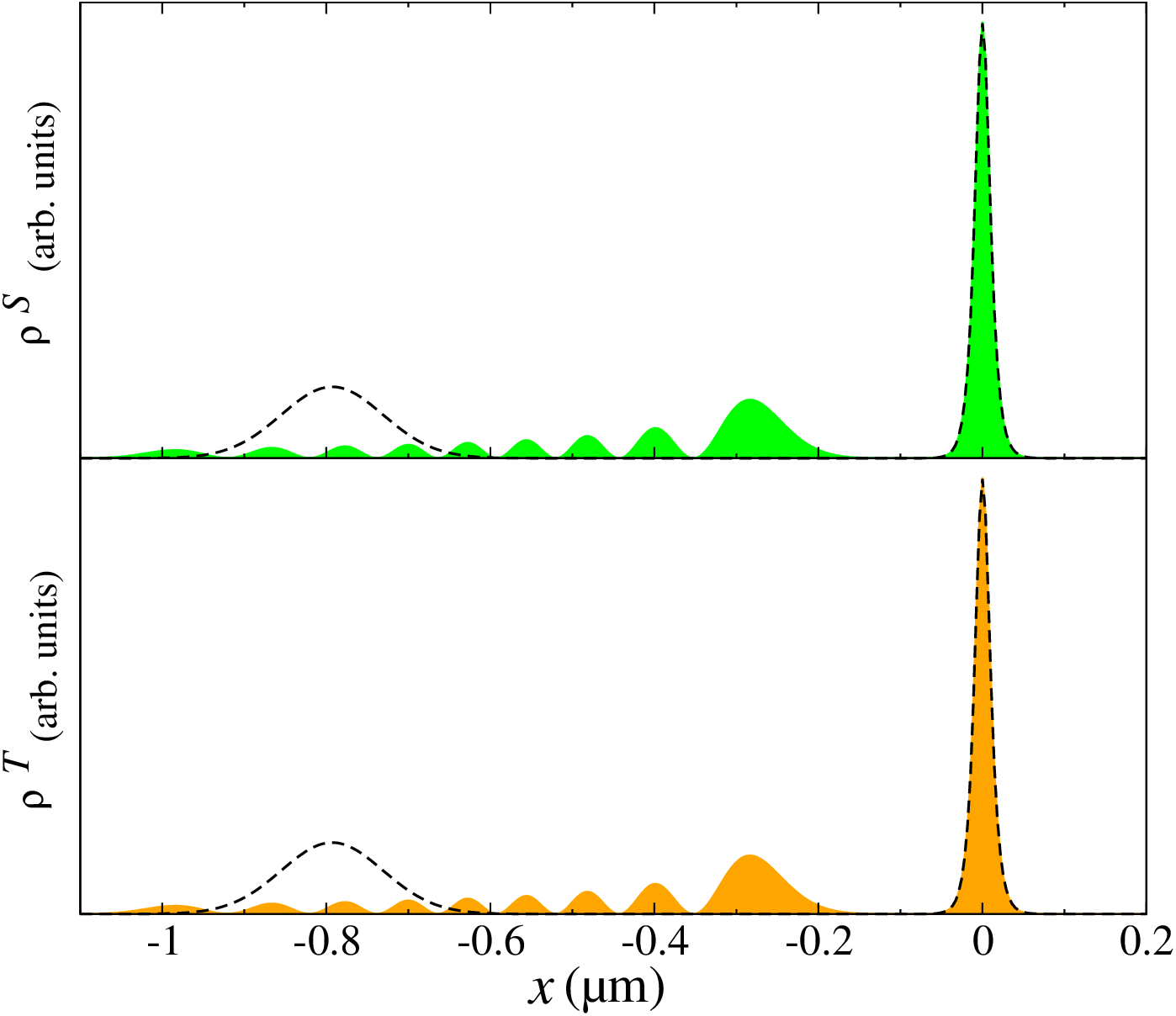}
\caption{(color online). Initial (dashed) and final electron distribution (full)
in arbitrary units when both singlet (top) and triplet (bottom)
are almost totally reflected.}
\end{center}
\end{figure}
\begin{figure}
\begin{center}
\includegraphics[width=7.5cm,height=4cm]{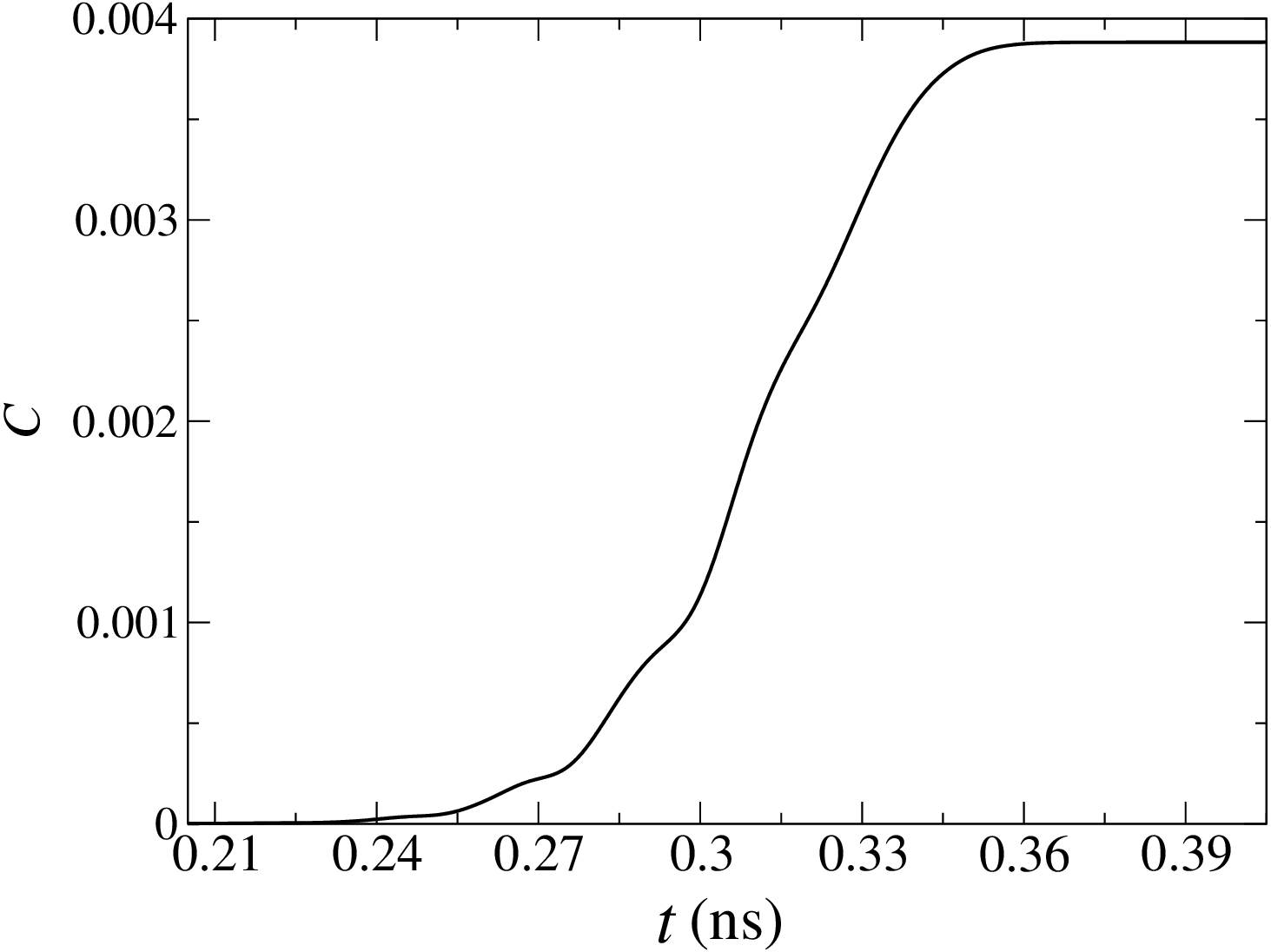}
\caption{(color online). Concurrence as a function of time when both singlet and
triplet are almost totally reflected.}
\end{center}
\end{figure}

A third regime of interest is when both singlet and triplet are
almost fully transmitted. Figure 9 shows the initial and final
electron density for such a case with parameters $V_{w}=70.5$ meV,
$l_{w}=10$ nm and $V_{o}=10$ meV. Although for this choice of
parameters the quantum well can bind two electrons in the absence
of the SAW, the second electron does not in fact become bound when
it is carried by the SAW and after scattering the probability of
finding both electrons in the well is negligible. This is because
the SAW period is too short for the second electron to become
trapped in the well. We thus see that both singlet and triplet
states are almost perfectly transmitted, whilst the electron in
the quantum well remains very well-localised. However, the
electron in the SAW potential minimum, after passing through the
region of the bound electron in the well, gains energy from the
interaction, resulting in a superposition state which includes
excited states of the SAW. It is easily verified that for this
special case the concurrence takes the form \cite{concur}
$C=|$Im$\langle T|S\rangle|$, where $|T\rangle$ and $|S\rangle$
are the single electron states of the transmitted electron in a
SAW minimum which result from triplet and and singlet states
respectively. We see from this formula that $C=1$ only when
$|T\rangle$ and $|S\rangle$ differ by a phase factor of $\pi/2$.
This is not the case in general, for which not only the phases but
also the amplitudes of $|T\rangle$ and $|S\rangle$ are different.
Limits which give zero concurrence are when $|S\rangle=|T\rangle$
(such as the trivial case of no interaction between the electrons)
and when $|S\rangle$ and $|T\rangle$ are orthogonal. As explained
in the next section the latter can occur, or approximately so,
when an electron in the well resonantly tunnels out of the well
into an excited state of a SAW minima for singlet but not triplet
or visa versa. More generally, the overlap (and hence $C$) may be
small but not precisely zero due again to different tunnelling
rates for singlet and triplet. Figure 10 shows how the concurrence
and the relative probabilities develop in time. Similarly with the
previous cases the concurrence increases and saturates to a
constant value, while for intermediate times it oscillates. Since
the reflected part is very small, we get the expected result that
the asymptotic value of the transmitted concurrence approximately
equals the total concurrence $C\approx C_{t}\sim0.53$.

\begin{figure}
\begin{center}
\includegraphics[width=7cm,height=6cm]{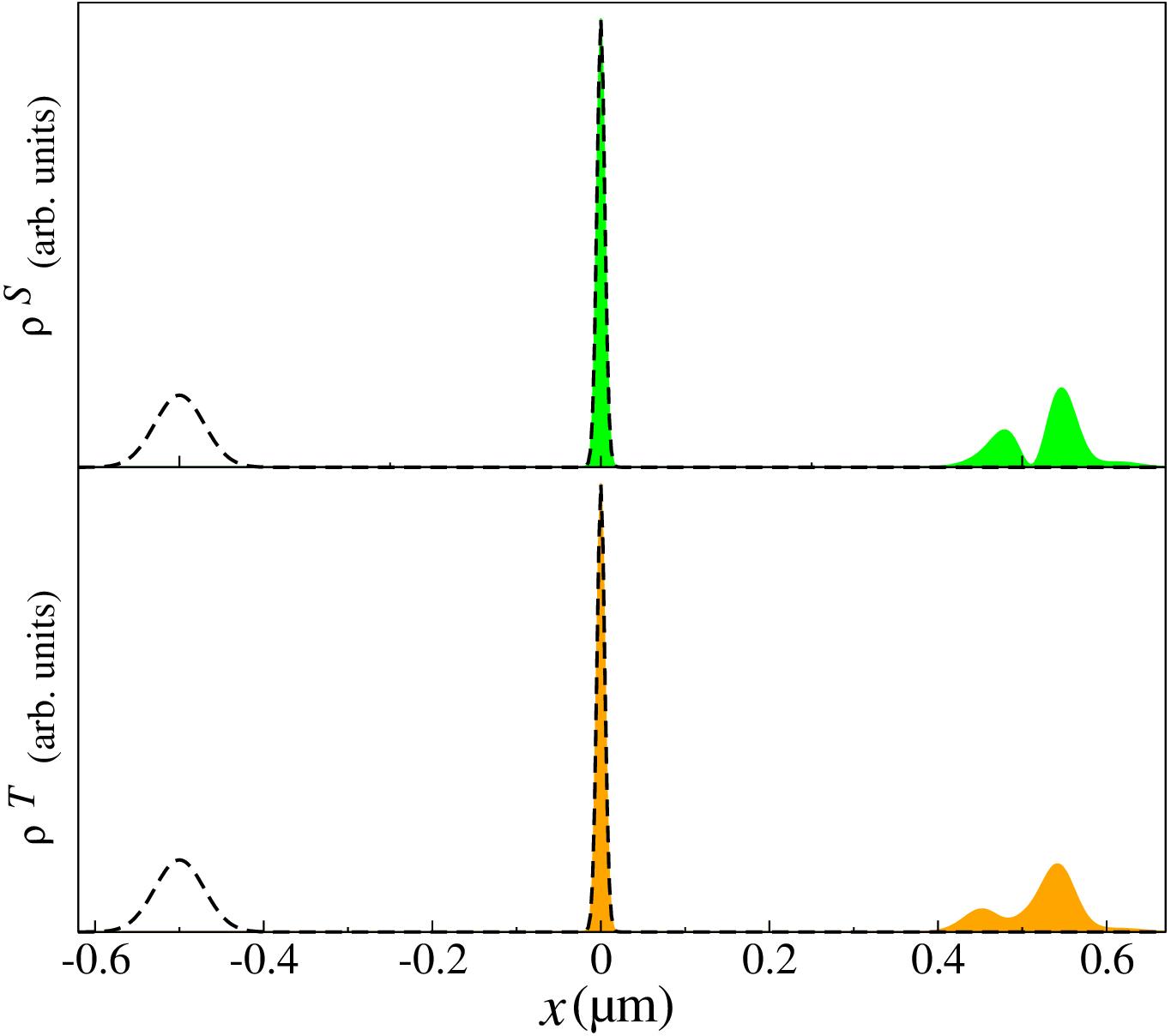}
\caption{(color online). Initial (dashed) and final electron distribution (full)
in arbitrary units for a typical case when both singlet (top) and
triplet (bottom) are almost fully transmitted.}
\end{center}
\end{figure}
\begin{figure}
\begin{center}
\includegraphics[width=7.5cm,height=8cm]{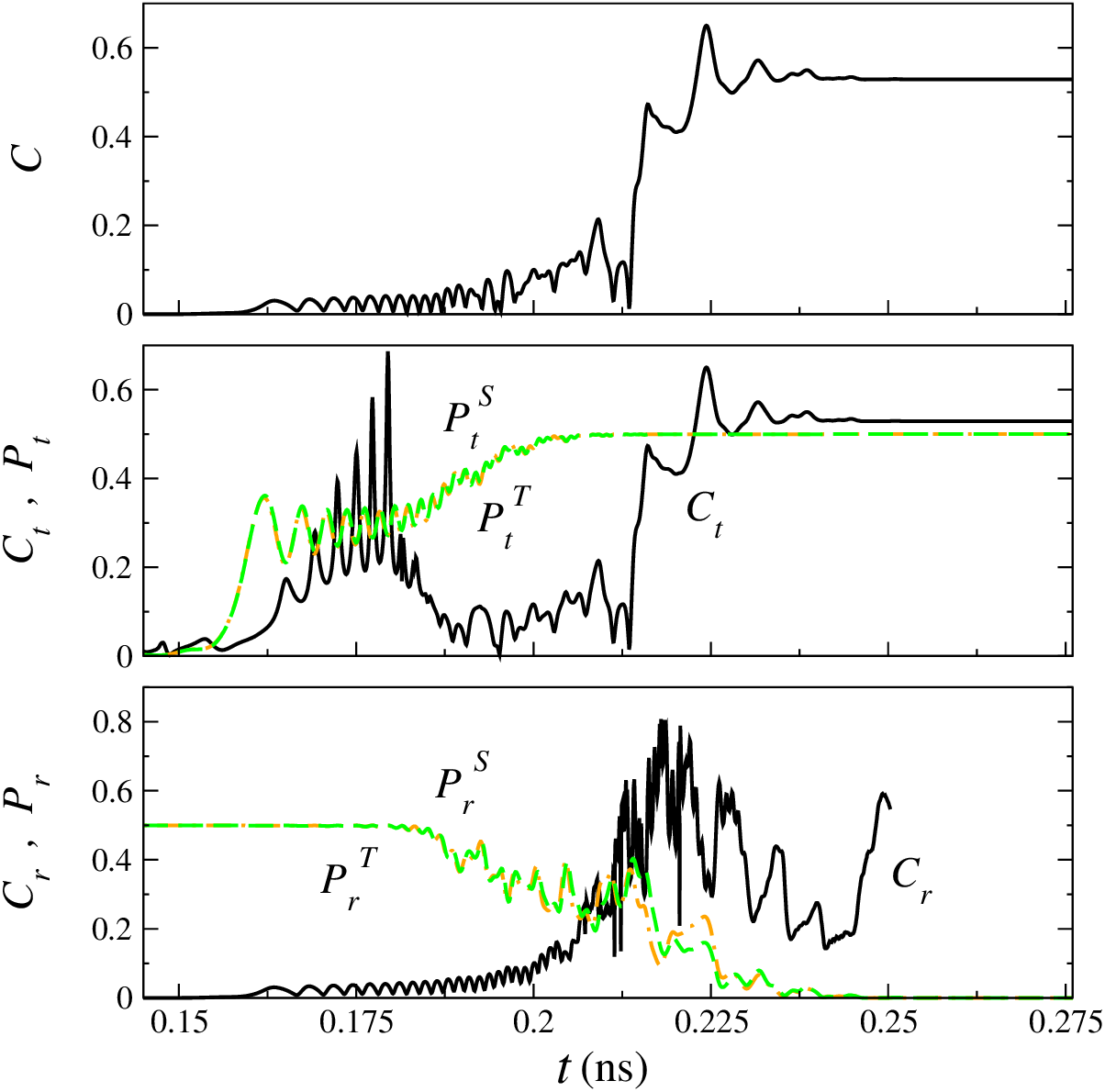}
\caption{(color online). Concurrence and relative probabilities as a function of
time for a typical case when both singlet and triplet
are almost fully transmitted.}
\end{center}
\end{figure}

Finally, we consider the possibility of choosing parameters such
that by changing the confining characteristics of the well the
singlet and triplet orbital components, after the scattering
event, may be chosen to differ only by a phase factor
$e^{i\delta\varphi}$, with
$\delta\varphi=\varphi_{S}-\varphi_{T}$. This may be done, at
least approximately, for parameters which give almost perfect
transmission for both singlet and triplet states. This occurs, for
example, when the SAW amplitude is sufficiently large. For this
regime the concurrence takes the form \cite{concur}
$C=|\sin\delta\varphi|$, as can be seen directly from Eq. (15),
using the form
$\Phi^{S}(x_{1},x_{2},t_{f})=e^{i\varphi_{S}}[\psi_{f}(x_{1})\varphi(x_{2})+\varphi(x_{1})\psi_{f}(x_{2})]/\sqrt{2}$,
$\Phi^{T}(x_{1},x_{2},t_{f})=e^{i\varphi_{T}}[\psi_{f}(x_{1})\varphi(x_{2})-\varphi(x_{1})\psi_{f}(x_{2})]/\sqrt{2}$,
for the singlet and triplet orbital components respectively. Note
that $\psi_{f}$ describes the electron in the SAW potential after
scattering and $\varphi$ describes the electron in the well. We
see immediately from this form that the concurrence may be
controlled by changing the relative phase of singlet and triplet
whilst maintaining approximately full transmission, giving full
entanglement when the magnitude of this phase difference is
$\pi/2$. However the value of the phase difference cannot be
easily controlled and, indeed, the phase difference picture is
itself only an approximate since the singlet and triplet
probabilities distributions in the SAW minimum after scattering
are never precisely identical. However for some special cases this
model is an excellent approximation as shown for example in Fig.
11 for the parameters $V_{w}=66$ meV, $l_{w}=10$ nm for which the
well can bind at least two electrons and $V_{o}=20$ meV. We see
that the electron in the SAW after the scattering event is
well-bound in the SAW potential minimum occupying the
characteristic ground state both for singlet and triplet. In Fig.
12 we show the concurrence as a function of time. In this case the
concurrence increases relatively smoothly compared to the previous
cases because we are in a regime where the one electron orbital
states in the scattering process are the same for singlet and
triplet, apart from a phase factor. Note that in order to even
have the possibility of achieving high concurrence the electrons
must have sufficient time to interact as the SAW propagates. The
timescale to give spin entanglement is of order $\hbar/|J|$, with
$J=E^T-E^S$ the exchange energy and $E^T$, $E^S$ the triplet and
singlet energies when we fix the SAW with both electrons in the
proximity of the well. Hence the SAW period has to be at least as
long as this for high concurrence to be possible and this is
indeed the case for typical SAWs which used to give high accuracy
single electron quantisation \cite{shilton,valery_1,valery_2}, as
we demonstrate in next section. Finally, it is worth mentioning
that to a good approximation a phase difference may be present
even when both states are reflected backwards, as described
earlier. However in this case the phase difference is expected
very small due to the weak effect of the Coulomb interaction.
Furthermore, the regime of near perfect transmission also has the
advantage that the electron in the SAW, after the scattering event
is very well-localised in a particular SAW minimum driven along
the wire.

\begin{figure}
\begin{center}
\includegraphics[width=7cm,height=6cm]{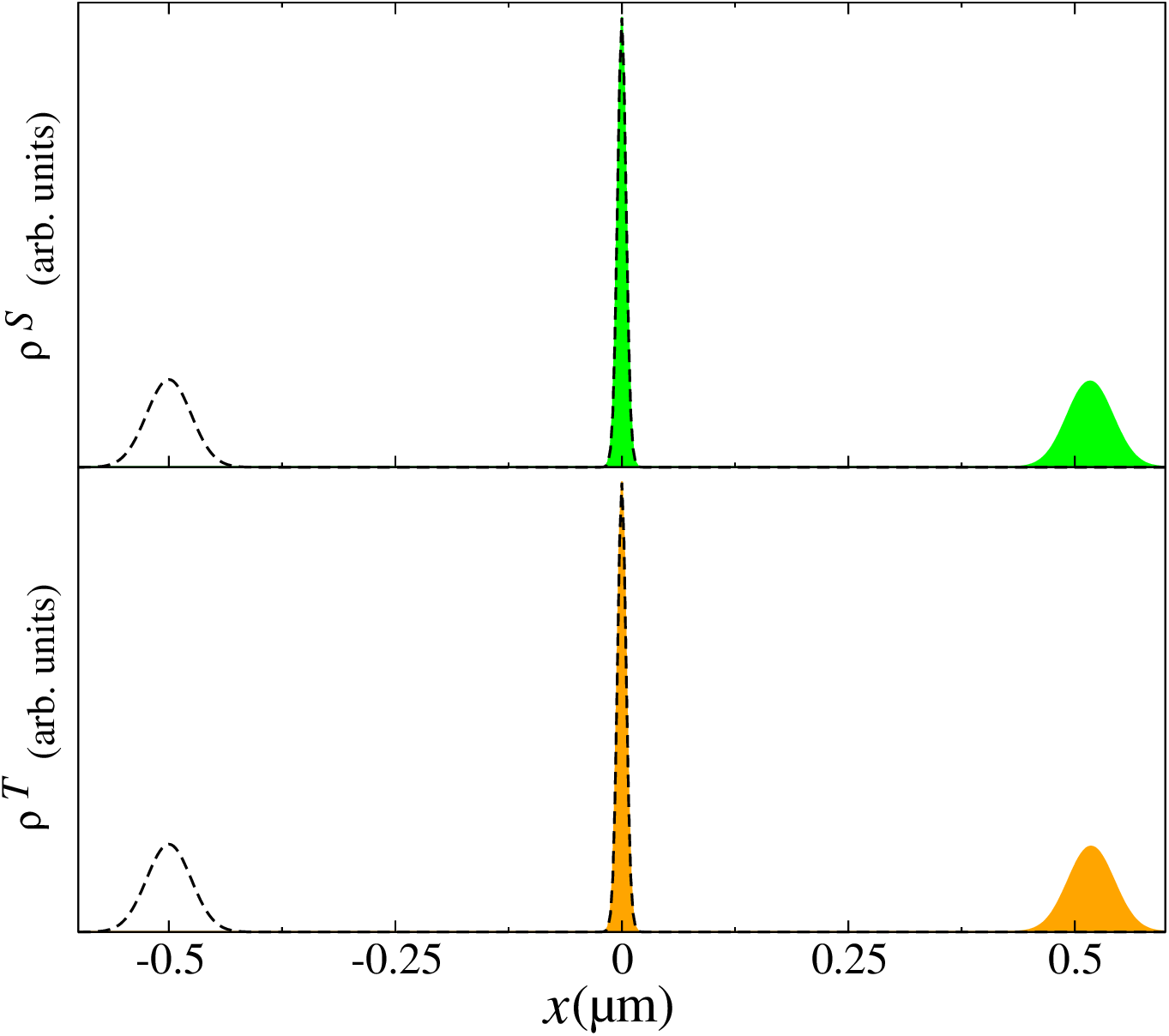}
\caption{(color online). Initial (dashed) and final electron distribution (full)
in arbitrary units when both singlet (top) and triplet (bottom)
are fully transmitted with a phase difference as it is defined in
the text.}
\end{center}
\end{figure}
\begin{figure}
\begin{center}
\includegraphics[width=7.5cm,height=8cm]{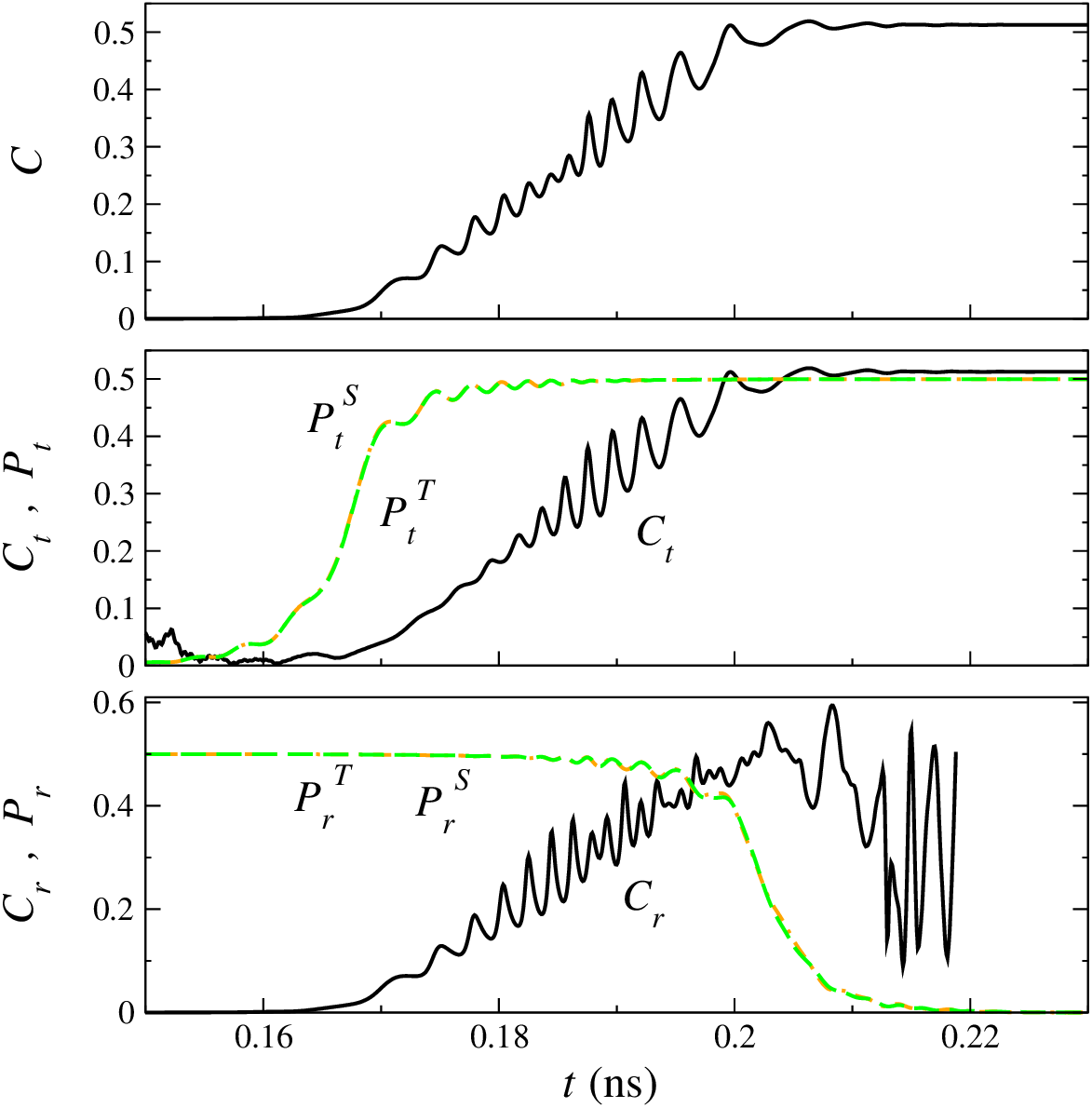}
\caption{(color online). Concurrence and relative probabilities as a function of
time when both singlet and triplet are fully transmitted with
a phase difference as it is defined in the text.}
\end{center}
\end{figure}

For all the cases we have described so far the induced entanglement between the two-electron spins is subjected to quantum decoherence which is an undesirable factor present to all solid state systems. The spin lifetime in GaAs, within which the process of generation-detection needs to take place, is estimated to be $\sim$100 ns \cite{kikkawa} arising primarily from phonon scattering. Typical times to generate entanglement in the SAW-based system are almost two orders of magnitude shorter while methods to read the final spin states have been described theoretically in Refs 11, 14 for the SAW-qubit and already demonstrated experimentally for the static qubit\cite{hanson}. Other important sources of decoherence that can affect the entanglement generation are coupling of electron spins to nuclear spins\cite{petta,loss}, noise on surface gates and temperature effects. These are discussed in the original proposal for SAW-based quantum computation\cite{barnes_4,furuta}.

\section{A Hartree approximation}

The mechanism that controls the different scattering of singlet
and triplet may be understood with an approximate treatment which
also gives insight into the origin of the differences in
transmission and reflection probabilities and concurrence. In a
mean-field approximation, the electron which is carried by the SAW
potential feels an effective time-dependent potential of the form
$V^{e}_{SAW}(x,t)\approx V_{t}(x,t)+V_{H}(x,t)$, where the second
term represents the Hartree potential due to the Coulomb repulsion
of the trapped electron in the well: $ V_{H}(x,t)=\int |\varphi
(x^{'},t)|^{2} V_{c}(x,x^{'}) dx^{'}$. This assumes that the
trapped electron in the quantum well remains well-localised and is
described at a specific time $t$ by the state $\varphi (x,t)$.
Below we explain the different scattering results of Sec. III.C by
employing the effective potential form $V^{e}_{SAW}$ for the
propagating electron.

\subsection{The single bound energy level regime}

First we consider the cases for which the quantum well has a
single bound energy level, i.e. the first two cases of Sec. III.C.
Figure 13(a) illustrates the effective potential
$V^{e}_{SAW}(x,t)$ when $t$ is such that $V_{SAW}(x,t)$ is minimum
at $x\sim0$ and for parameters that result in high transmission
for the singlet state and high reflection for the triplet (that is
the first case that we described in Sec. III.C). The effective
potential may be described as a triple well structure that changes
with time due to the time-dependent nature of the SAW potential.
Specifically, due to the SAW propagation, the right well becomes
deeper than the left well with increasing time, with the middle
well shifting upwards and downwards in energy at $x\sim0$.
Initially, the electron that is carried by the SAW resides in the
left well and has a tendency to tunnel through the middle well
into the right well, in order to remain bound in the SAW
potential. The tunnelling mechanism is more efficient when there
are resonance conditions for the electron to first tunnel from the
left well into the middle well and then from the middle well to
the right-hand well, i.e. in the time interval when resonant bound
state energy widths of the left, middle and right-hand wells
overlap. Of course the SAW potential amplitude, the width of the
wire and the characteristic width and depth of the well should be
chosen in such a way that the resulting effective potential
guarantees at least one resonance energy level for the middle
well, that will lie above the bottom of the right and left wells.
A sufficiently large SAW amplitude is also necessary to ensure
that the barrier between the resonance condition between left and
middle wells is satisfied, otherwise the electron in the SAW will
be reflected. Finally, a necessary criterion for high transmission
is that there must be sufficient time for the whole process to
take place, i.e. the tunnelling time into and out of the middle
well must be much smaller than the period of the SAW, a condition
that is fulfilled in the simulations.

\begin{figure}
\begin{center}
\includegraphics[width=7.5cm,height=8.5cm]{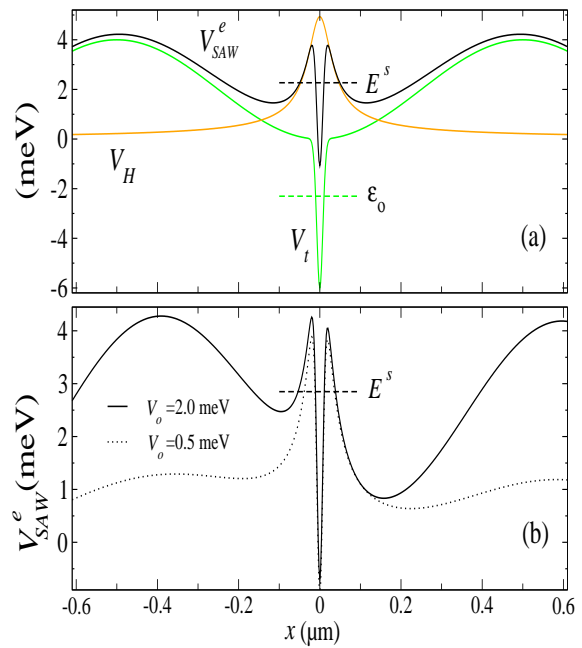}
\caption{(color online). (a) The effective potential, and its constituent parts,
that a SAW electron feels at a time for which the SAW potential is
minimum at $x=0$ and for a SAW amplitude $V_{o}=2$ meV. (b) The
effective potential close to the resonant tunnelling regime for
the singlet state for a SAW potential amplitude $V_{o}$=2 meV
(solid line). For $V_{o}$=0.5 meV (dotted line) the resonance
condition can not be fulfilled (see text).}
\end{center}
\end{figure}

To explain qualitatively the difference in the evolution between
the singlet and the triplet states, we also need to consider
explicitly the symmetry of the orbital states and take into
account the fact that only for the singlet state can both
electrons occupy the same one electron orbital state. More
specifically, if $\varphi_{o}(x,t)$ is the lowest resonant bound
state of the combined well and SAW potential that peaks in the
region of the well $x\sim0$, then the instantaneous energy of the
two-electron singlet state on resonance is $E^{S}(t)\approx
\varepsilon_{o}(t)+U_{o}(t)$ where $ U_{o}(t)=\int
|\varphi_{o}(x_{1},t)|^{2}
V_{c}(x_{1},x_{2})|\varphi_{o}(x_{2},t)|^{2}dx_{1}dx_{2}$ is the
Coulomb energy when both electrons occupy the single electron
state $\varphi_{o}(x,t)$. This is analogous to resonant tunnelling
in the Anderson impurity model. A similar approach could be
applied to the triplet state but in this case the two electrons
must occupy different one electron resonance levels,
$\varphi_{o}(x,t)$ and $\varphi_{1}(x,t)$ due to the Pauli
principle. If the quantum well had a second resonant state then
the two-electron resonance would occur at the generally higher,
triplet resonance energy $E^{T}(t)\approx
\varepsilon_{1}(t)+U_{1,0}(t)-J_{1,0}(t)>E^{S}(t)$, where
$U_{1,0}(t)$ and $J_{1,0}(t)$ are the Coulomb and the exchange
integrals respectively.

From the above description it is clear that the electron which is
carried by the SAW feels an effective potential which is
independent of the character of the two-electron orbital state
(symmetric or antisymmetric), however this is not the case for the
energy levels of the tunnelling process. For the first regime
described in Sec. III.C the singlet resonance level gives high
resonance transmission only for the singlet state, as expected.
The transmission of the triplet state is much weaker and is in
fact due to non-resonant tunnelling, since with the chosen
parameters the energy of the electron in the effective potential
$V^{e}_{SAW}$ is always below the barriers which define the middle
well. Note that for this case the SAW potential amplitude is
strong enough to drive the propagating electron to the resonance
level. On the other hand, in the second regime described in Sec.
III.C the SAW potential well is so shallow that, for both singlet
and triplet, the propagating electron is reflected before it
reaches the resonance energy for tunneling into the middle well.
Figure 13(b) shows the effective potential profile close to the
resonant tunnelling regime for the singlet state for both SAW
potential amplitudes. Note that the resonance level lies above the
bottom of the left and right wells, ensuring that tunnelling may
occur. For these two regimes an effective antiferromagnetic
exchange interaction controls the scattering process since the
singlet scattering involves lower energy levels than the triplet.

\subsection{Beyond the single bound energy level regime}

For the last two cases of Sec. III.C the quantum well has more
than a single bound energy level and can bind at least two
electrons. One effect of making the quantum well deeper is to
reduce the barriers to the SAW wells to the left and right, as can
be seen by comparing Figs. 13(a) and 14(c) for the effective one
electron potential. This results in almost perfect transmission
for both singlet and triplet states provided the SAW amplitude is
much larger than the small residual barriers when the quantum well
is at a SAW potential minimum (Fig. 14(c)). However, the different
positions of the singlet and triplet resonances still affect the
final orbital states of the transmitted electron in the SAW,
depending on the magnitude of the tunnel barrier when the SAW
potential minimum energy is close to the resonance level in the
quantum well (e.g. Fig. 14(b)).

Specifically, when this barrier is large the propagating electron
emerges in the lowest state of the SAW potential minimum. This is
illustrated in Fig. 14, where we plot some of the instantaneous
eigenenergies of the effective one electron potential
$V^{e}_{SAW}$ for the parameters that result in a phase difference
between singlet and triplet (this is the last case that we
considered in Sec. III.C). Note that the quasi-bound state levels
within the well include the effect of Coulomb repulsion due to the
bound electron which shifts the potential well up in energy by
$V_{H}$ and also gives rise to the very small peaks in the
effective potential. At $t=0$ (Fig. 14(a)) the propagating
electron is in the lowest energy of the $V^{e}_{SAW}$ potential
minimum to the left of the quantum well. This is actually the
first excited state of the system since the lowest state is in the
well. Between $t=0$ and $t=0.3T$ (Figs. 14(a),(b)) the energy
levels corresponding to the electron in the $V^{e}_{SAW}$ minimum
and in the second state of well are almost the same (anti-crossing
region) but there is insufficient time for the electron to tunnel
into the well and therefore it remains in the $V^{e}_{SAW}$
potential minimum. It therefore makes a (non-adiabatic
Landau-Zener) transition from the first to the second excited
state of the system. At $t\simeq0.3T$ there is a further
anti-crossing region and transition to the third excited state of
the system with the electron remaining in the SAW. Between
$t=0.5T$ and $t=T$ (Figs. 14(d),(e)) there are further transitions
back to the initial state. This is the reason that the propagating
electron emerges in the lowest state of the $V^{e}_{SAW}$
potential minimum which actually coincides with the original SAW
potential minimum when the electrons are well separated at $t=T$
(Fig. 14(e)) when the SAW cycle is completed. It is clear from
Fig. 14(c) that the highest resonance level of the well, in this
case is the fourth level, gives rise to the largest interaction
with the propagating electron as long as tunnelling to lower
excited states is blocked due to the large barrier. Of course
lower excited resonances are involved for shallower quantum wells.

Although the scattering process does not excite the electron in
the SAW, it does affect its wave function by inducing a phase
shift and this phase shift is different for singlet and triplet
cases due to Coulomb interaction. In particular, the evolution of
singlet and triplet states will be different but, unlike the
lowest singlet-triplet pair, the higher-lying levels will
generally have the triplet lower in energy than the singlet, due
essentially to Hund's rule \cite{denver,bryant}. This results in a
ferromagnetic exchange interaction, rather than the
antiferromagnetic exchange of the lowest singlet-triplet pair. It
is the small energy difference between the relevant singlet and
triplet energy levels which induces the relative phase between
singlet and triplet states as a consequence of the interaction
time for the two electrons which is set by the SAW period. We show
in the next section how the phase difference may be directly
related to a ferromagnetic exchange interaction between the spins
of the two electrons as they interact, changing their
entanglement.

Finally, when the parameters are such that the probability to
tunnel from the SAW potential minimum to the well is not
negligible in the anti-crossing region, (e.g. the third case
considered in Sec. III.C) then the electron will emerge in a
superposition state of the low-lying states of the SAW. This can
be understood qualitatively again by referring to Fig. 14. Since
the tunnelling probability is no longer negligible at the point
where the $V^{e}_{SAW}$ minimum crosses a resonance level, then
the electron in the region of the well (e.g. Fig. 14(c)) will
emerge in a superposition state of the third and fourth energy
levels. Similarly, at later times when the energy level
corresponding to the electron in the well sweeps through higher
excited levels in the $V^{e}_{SAW}$ potential minimum the electron
will eventually leave the well and emerge in an asymptotic state
that is a superposition of the low-lying states of the SAW. In
this regime the different orbital states for singlet and triplet
are due to different tunnel barriers for the highest-lying
singlet-triplet pair due to different positions of the resonances.

\begin{figure}
\begin{center}
\includegraphics[width=7.5cm,height=10.5cm]{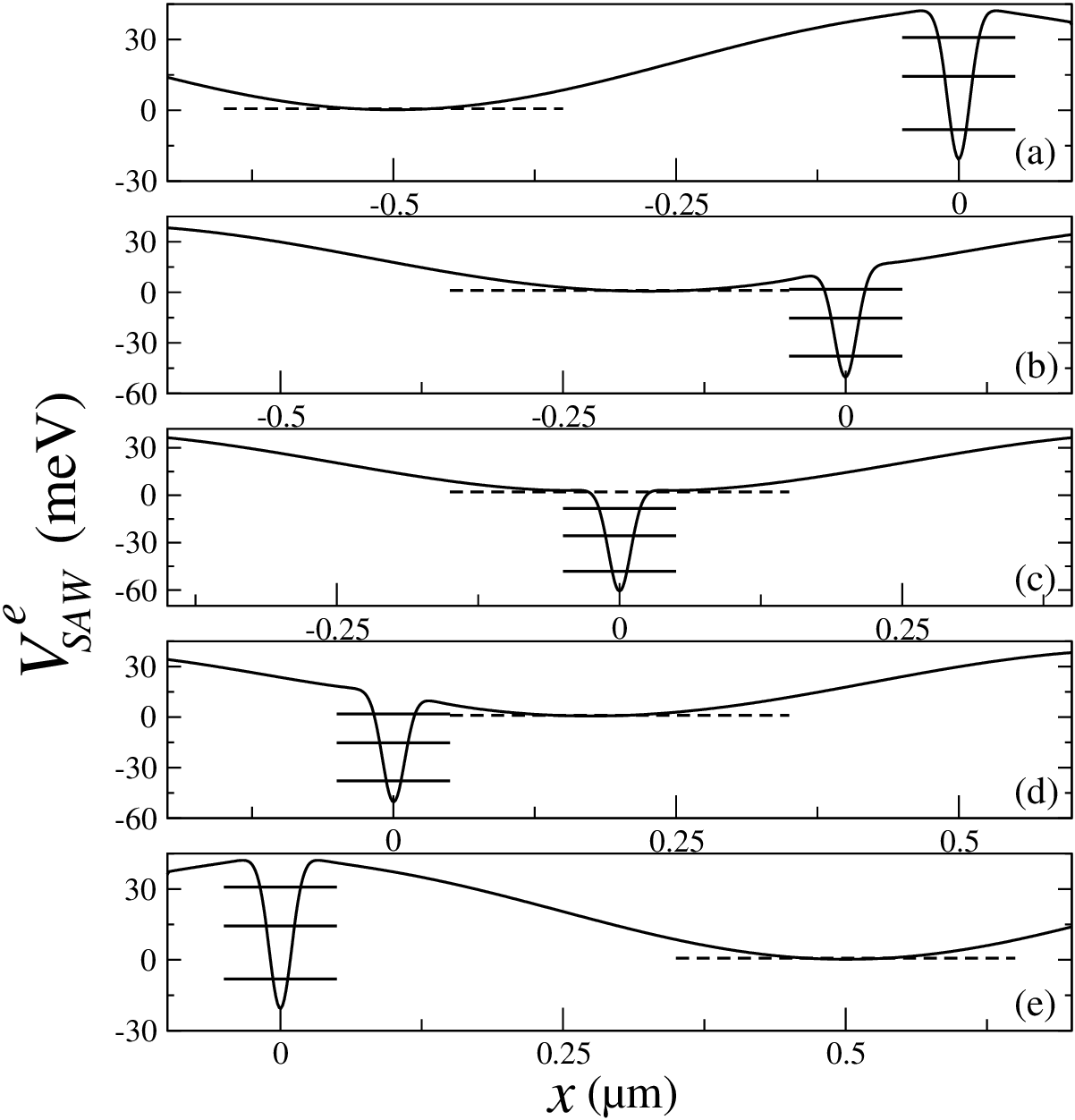}
\caption{Propagation of an electron in the $V^{e}_{SAW}$ minimum
passing through the well region. The dashed line indicates the
energy of the propagating electron at each particular time, if
after the SAW cycle it exits the well region in the lowest state
of the SAW minimum. The full lines indicate the energy levels of
the quantum well. In the anti-crossing regions (b),(d) the
electron makes a non-adiabatic Landau-Zener transition and always
remains in the $V^{e}_{SAW}$ minimum when the tunnel barrier is
large and as a result after the SAW cycle the electron emerges in
the lowest state of the SAW potential minimum (e). When the
tunnelling probability into and out of the quantum well is not
negligible, the electron emerges in a superposition state of the
low-lying states of the SAW. The time sequence is from (a) to (e)
and specifically $t/T$=0, 0.3, 0.5, 0.7, 1.}
\end{center}
\end{figure}

To conclude, an effective antiferromagnetic exchange interaction
controls the scattering events when the quantum well has only one
bound state, due to the singlet resonance channel. However, by
increasing the depth of the well the ground singlet resonance
level becomes inactive simply because it lies much lower than the
energy of the propagating electron at all times. In this regime
the scattering is controlled by an effective ferromagnetic
exchange interaction involving excited states for singlet and
triplet in which the triplet is lower. It is interesting to note
that in the flying qubit scheme \cite{barnes_4,gumbs} it is always
an antiferromagnetic exchange interaction that generates the
entanglement, whereas in the scheme that we propose both
ferromagnetic and antiferromagnetic type interactions can generate
entanglement depending on SAW and well parameters.

\section{Some general features of the entanglement}

In this section we generalise some of the above results and
demonstrate quantitatively the sensitivity of the system to
changes in the well and the SAW characteristics.

\begin{figure}
\begin{center}
\includegraphics[width=8cm,height=8cm]{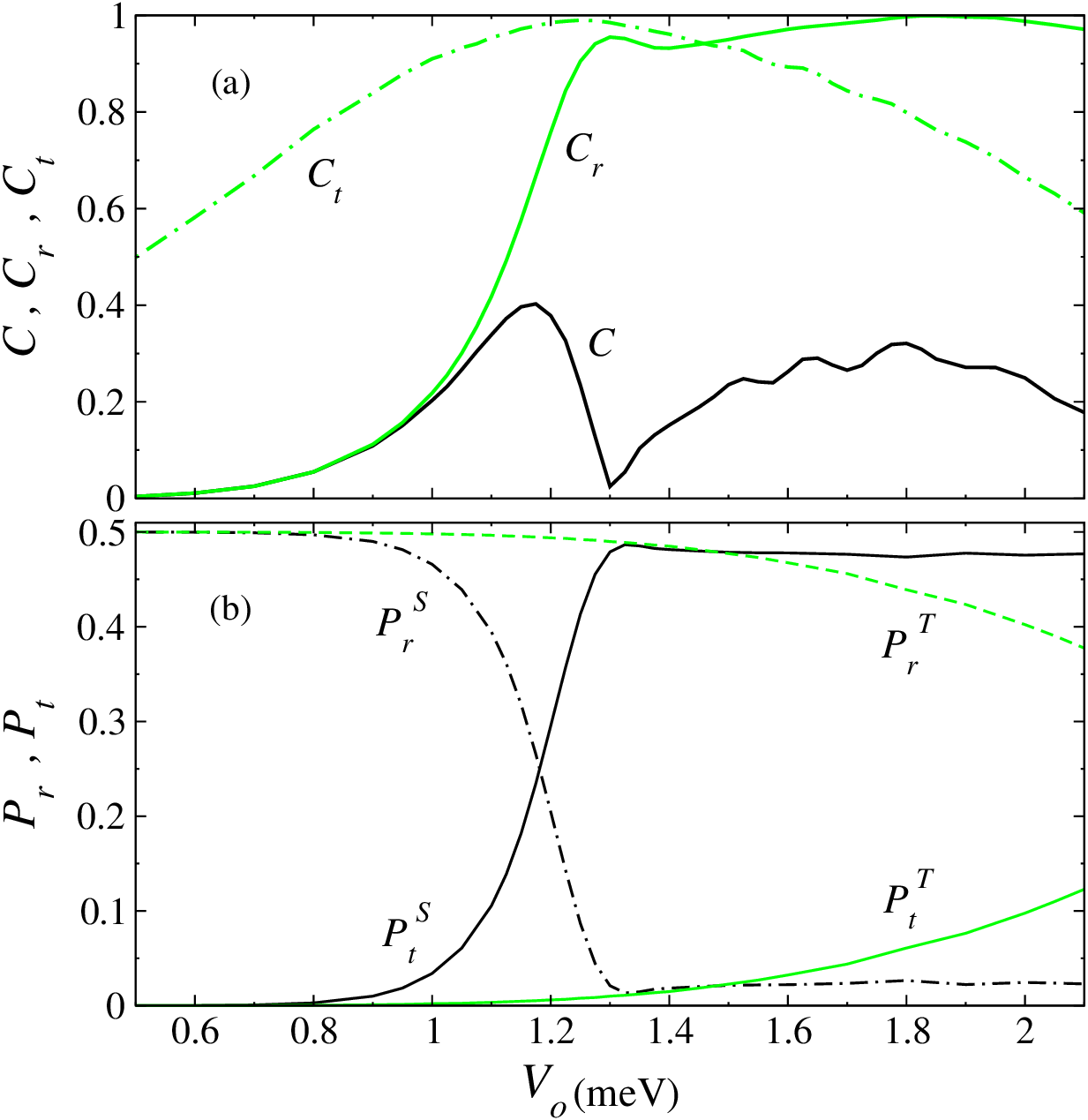}
\caption{(color online). Variation of asymptotic (final time) concurrence (a) and
relative probabilities (b), as defined in the text, versus the SAW
potential amplitude when the quantum well is such that only a
single electron can be bound.}
\end{center}
\end{figure}

Figure 15(a) illustrates the variation of the concurrence versus
the SAW potential amplitude for a quantum well with $V_{w}=6$ meV
and $l_{w}=7.5$ nm, which can accommodate only a single bound
electron. This plot shows the total, transmitted and reflected
concurrence at the final time for which the overlap of propagating
and bound electron wave packets is negligible. Figure 15(b)
presents the corresponding probabilities. Note that the SAW
potential amplitude is restricted to the specific regime for which
the electron in the quantum well remains well-localised, as
described in Sec. II B. We see that the very small transmission of
singlet and triplet (which we include here for completeness),
corresponding to the minimum value of the SAW amplitude, gives
rise to a concurrence, $C_{t}\sim0.5$. With increasing SAW
amplitude there is then a regime in which the transmitted
concurrence increases to $C_{t}\sim1$, for which the singlet state
is on resonance and simultaneously there is minimum transmission
for the triplet state. We may regard this as a two-electron spin
filter for which the initial unentangled state, that is an equal
superposition of singlet and $S_{z}=0$ triplet states, has its
triplet component filtered out (reflected) with resonant
transmission of the fully entangled singlet component. For this
SAW amplitude, a ``fuzzy'' position measurement applied to the
outgoing electron (say through charge sensing\cite{delft1,delft2,delft3}), which merely
resolves whether it is transmitted or reflected, could be used to
probabilistically prepare a highly entangled state. The form of
the state prepared is heralded by the measurement outcome. Further
increase of the SAW potential amplitude from this point causes the
transmitted concurrence to gradually decrease due simply to the
higher transmission of the triplet state. The reflected
concurrence is very small for low SAW potential amplitude and
approximately equals the total concurrence due to the very high
reflection for both states. It then increases smoothly as the
singlet transmitted part increases and remains almost constant and
close to unity with further increase of the SAW amplitude, since
the reflected component is mainly triplet. Finally, the total
concurrence has a relatively more complicated behavior, although
it is clear that it has a maximum value $C\sim0.4$ when
$P^{S}_{r}=P^{S}_{t}\sim0.25$, namely when the singlet is equally
transmitted and reflected. Also it is always lower than the
transmitted and reflected concurrence except when $P_{t}\sim0$ and
then $C\approx C_{r}$. We may conclude from Fig. 15 that the
degree of entanglement for the two electrons can be changed
significantly with SAW amplitude in the regime that scans through
the singlet resonance and that there exists a point where in
principle, through ``fuzzy'' position measurement, highly
entangled states could be prepared.

\begin{figure}
\begin{center}
\includegraphics[width=8.5cm,height=6.5cm]{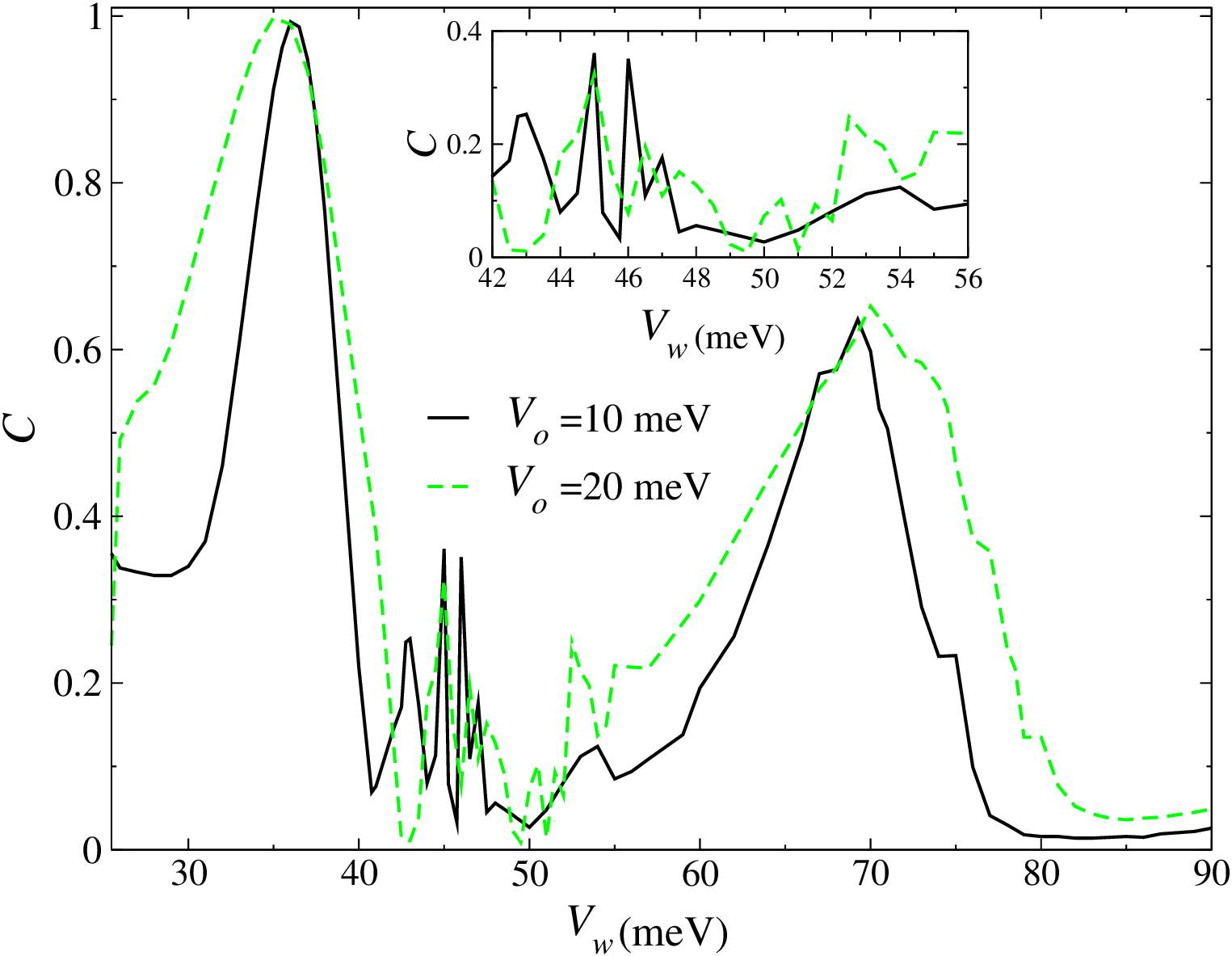}
\caption{(color online). Variation of asymptotic (final time) concurrence for a
fixed SAW potential amplitude ($V_{o}=10$ meV, $V_{o}=20$ meV) as
a function of the well depth.}
\end{center}
\end{figure}

A more relevant case for simpler experiments, which don't require
position measurement to project into the transmitted or reflected
outcomes for the outgoing electron, is when the SAW potential
amplitude and the quantum well are such that the electron in the
SAW is always fully transmitted, or approximately so, leaving the
partner electron bound in the quantum well. The backward
reflection, which is more likely to occur for a low potential
amplitude, may cause undesirable effects, since the reflected
electron will occupy multiple wells and involves highly excited
components. This case of a reflected non-bound electron is more
efficiently studied using kinetic injection without the presence
of the SAW potential, as it is described for example in Refs. 19,
21. In addition, a strong SAW potential amplitude has the
advantage of preventing the trapped electron from leaking into
adjacent minima, thus minimizing possible errors. We have
calculated the concurrence as a function of the well depth for two
different, though relatively strong, SAW potential amplitudes of
$V_{o}=20$ meV and $V_{o}=10$ meV, and for fixed $l_{w}=10$ nm.
The SAW potential amplitude that is used in the experiments for
SAW-based SET applications can be even stronger than this
($V_{o}\sim40$ meV) \cite{robinson}, though along the channel
there is likely to be some screening due to the gate bias. The
chosen parameters guarantee that there is very high transmission
both for singlet and triplet states ($P^{S/T}_{t}\sim0.5$). In
this study the range of the well depth ensures high localisation
of the trapped electron resulting in a truly bound singlet ground
state, as calculated within a Hartree approximation and therefore,
a ferromagnetic type exchange interaction generates the
entanglement as described in the previous section.

The results for the total concurrence, which almost equals the
transmitted concurrence, are shown in Fig. 16, while Fig. 17
illustrates the singlet and triplet components of the electron in
the SAW (after scattering) for various well depths and for the SAW
amplitude of $V_{o}=10$ meV. Figure 16 presents two distinct
maxima for each of the two amplitudes considered with an
intermediate region of relatively low concurrence (which is shown
in detail in the inset of Fig. 16). Analysis of the data shows
that in the rise up to the first maximum, the asymptotic state is
approximated well by the simple phase difference picture described
in section III.C, i.e. with the electron in the SAW potential
minimum being in its ground state, to a good approximation, but
with a phase difference between singlet and triplet components.
This concurrence of almost unity at the maximum then corresponds
to a phase difference of $\delta\varphi \sim \pi/2$. As the well
depth is increased from this point the electron in the SAW
occupies additional excited states which are different for singlet
and triplet (the phase difference picture is no longer valid) and
this is why the concurrence decreases. Figure 17 helps us
understand how the singlet and triplet components of the electron
in the SAW change with well depth and specifically how we pass
from a region of different probability distribution to a region
where the phase difference picture is valid. This behavior is
clear for example by considering Figs. 17(a),(b) and (c). Similar
behavior is valid for a SAW amplitude of $V_{o}=20$ meV. Within
the intermediate region for both SAW amplitudes the concurrence
fluctuates due to spin-dependent scattering events which involve
excited states of the SAW potential minimum. Figure 17(c) shows an
example within this region. Note that zero concurrence corresponds
to cases where the orbital states in the SAW for singlet and
triplet components are exactly orthogonal. Further increase of the
well depth gives rise to a second concurrence maximum, due to the
fact that the final states of the SAW electron for singlet and
triplet components are approximated by the ground state of the SAW
potential minimum. Similar to the first concurrence maximum, the
phase difference picture is again valid as shown for example in
Fig. 17(d). Excited states with high probability however are
involved in the scattering process, different for singlet and
triplet as we demonstrate in Fig. 17(e) (reflecting spin-dependent
scattering) lowering the concurrence. This occurs up to the regime
of the very deep quantum well (at the right-hand side of Fig. 16
and for each SAW amplitude) where the phase difference picture
becomes valid as Fig. 17(f) demonstrates. This is because the
Coulomb interaction is effectively reduced due to the high
confinement of the trapped electron. Singlet and triplet
components are scattered mainly due to the presence of the
potential well in the same final states with only a small effect
from the Coulomb interaction, which will become negligible for
extremely deep wells. When this extreme limit is reached the
two-electron states will be approximated at all times by single
electron states.

\begin{figure}
\begin{center}
\includegraphics[width=8cm,height=8cm]{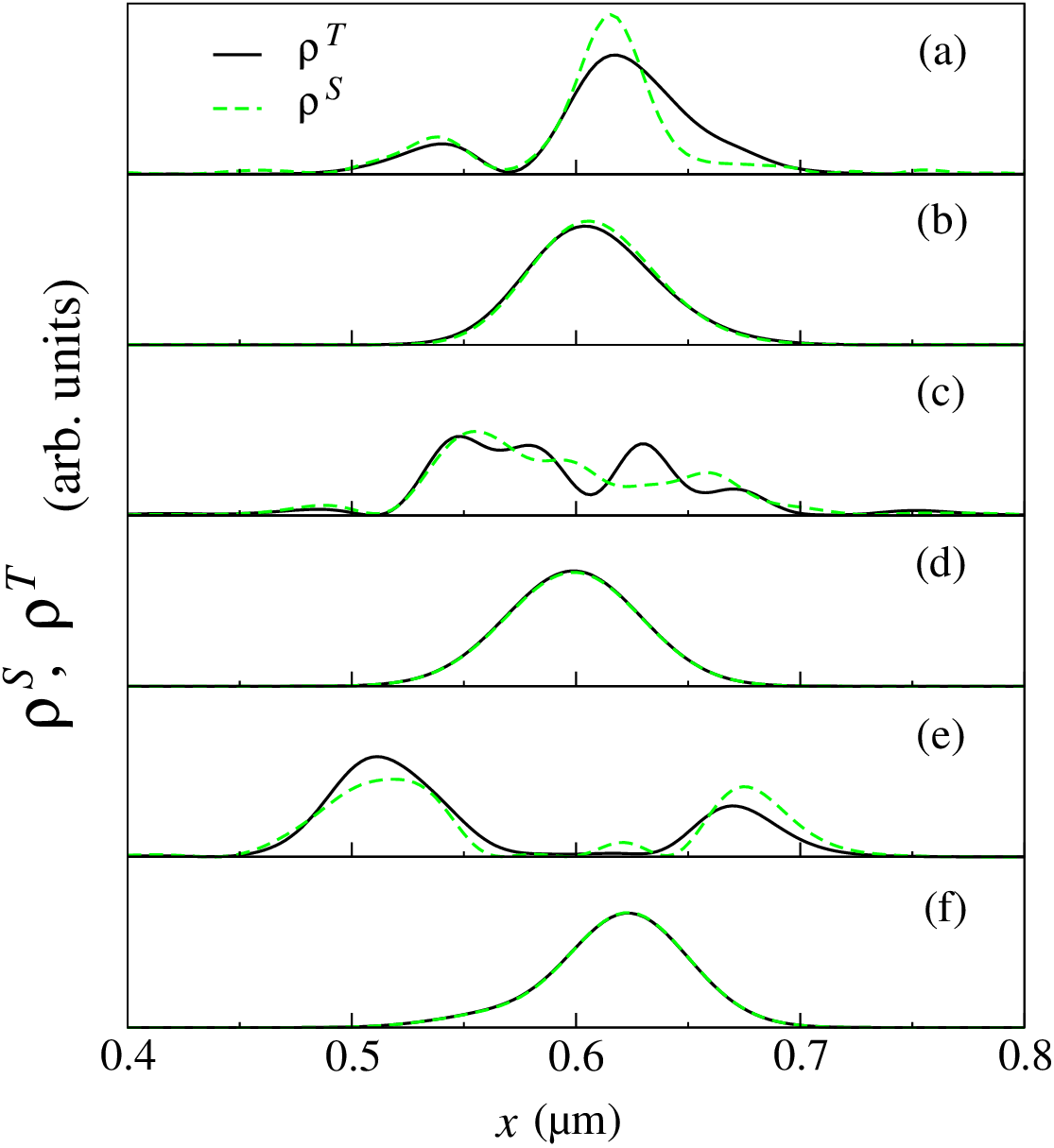}
\caption{(color online). Final electron distribution for singlet and triplet
states in the SAW potential minimum, for a SAW amplitude of
$V_{o}=10$ meV and a well depth from (a) to (f) of $V_{w}$=26, 36,
47, 62, 75, 85 meV.}
\end{center}
\end{figure}

As we have said in previous sections, the asymptotic value of the
concurrence (at the final time) which emerges when a relative
phase difference is present between singlet and triplet states,
depends on the magnitude of the so-called exchange energy
$J(t)=E^T(t)-E^S(t)$ and the SAW period which sets the interaction
time. In the phase difference regime an approximate Heisenberg
Hamiltonian \cite{mermin} $H(t)=J(t)\textbf{S}_{1}\cdot
\textbf{S}_{2}$, with $\textbf{S}_{\imath}$ the spin operator of
the $\imath$th electron can provide insight into the spin
entanglement generation. This is because in this regime the two
electrons at all times occupy different and well-defined orbital
states and as we have described in Sec. IV.B these states are the
same for singlet and triplet apart from a phase factor. The
exchange energy as a function of time can be determined by solving
the instantaneous (time-independent) two-electron Schr\"odinger
equation for singlet and triplet states treating time as a
parameter i.e.
$H(t)\Phi^{S/T}_{n}(t)=E^{S/T}_{n}(t)\Phi^{S/T}_{n}(t)$, with the
Hamiltonian given by Eq. (11). A common diagonalisation procedure
is described in Refs. 13, 41, 42. The instantaneous solutions
provide the sets $E^{S}_{n}(t)$ and $E^{T}_{n}(t)$ with $n$ the
eigenvalue index. By following the non-adiabatic Landau-Zener
transitions which successfully take place in the phase difference
regime, we can extract the energy of the two electrons during the
scattering event i.e. $E^T(t)$, $E^S(t)$ and from these the
$J(t)=E^T(t)-E^S(t)$ curve. In Fig. 18 we show the exchange energy
as a function of time for two different well depths $V_{w}$=36
meV, $V_{w}$=62 meV and a SAW potential amplitude $V_{o}$=10 meV
which result in a phase difference as shown in Figs. 17(b),(d). As
we have analysed in Sec. IV.B and we see in Fig. 18, the exchange
energy $J$ is negative in the phase difference regime. The lower
$J$ for the case of the $V_{w}=$62 meV well depth is because a
higher excited energy level for the singlet-triplet pair is
involved in the scattering process, compared to the $V_{w}=$36 meV
case and in general higher excited energy levels have a smaller
separation \cite{denver,bryant}. For the phase difference regime
and within the Heisenberg model we can calculate the asymptotic
concurrence $C=|\sin\delta\varphi|$ by extracting the relative
phase difference $\delta\varphi$ directly from the $J(t)$ curve as
$\delta\varphi=\int^{T}_{0} J(t)/\hbar dt$. Note that the time
interval of the integration is set by the SAW period $T$ which is
fixed in the experiments. The values that we take by this
approximate treatment are in excellent agreement with the values
that we take by solving the two-electron Schr\"odinger equation and
by calculating the concurrence by the original formula (15).

\begin{figure}
\begin{center}
\includegraphics[width=8cm,height=4cm]{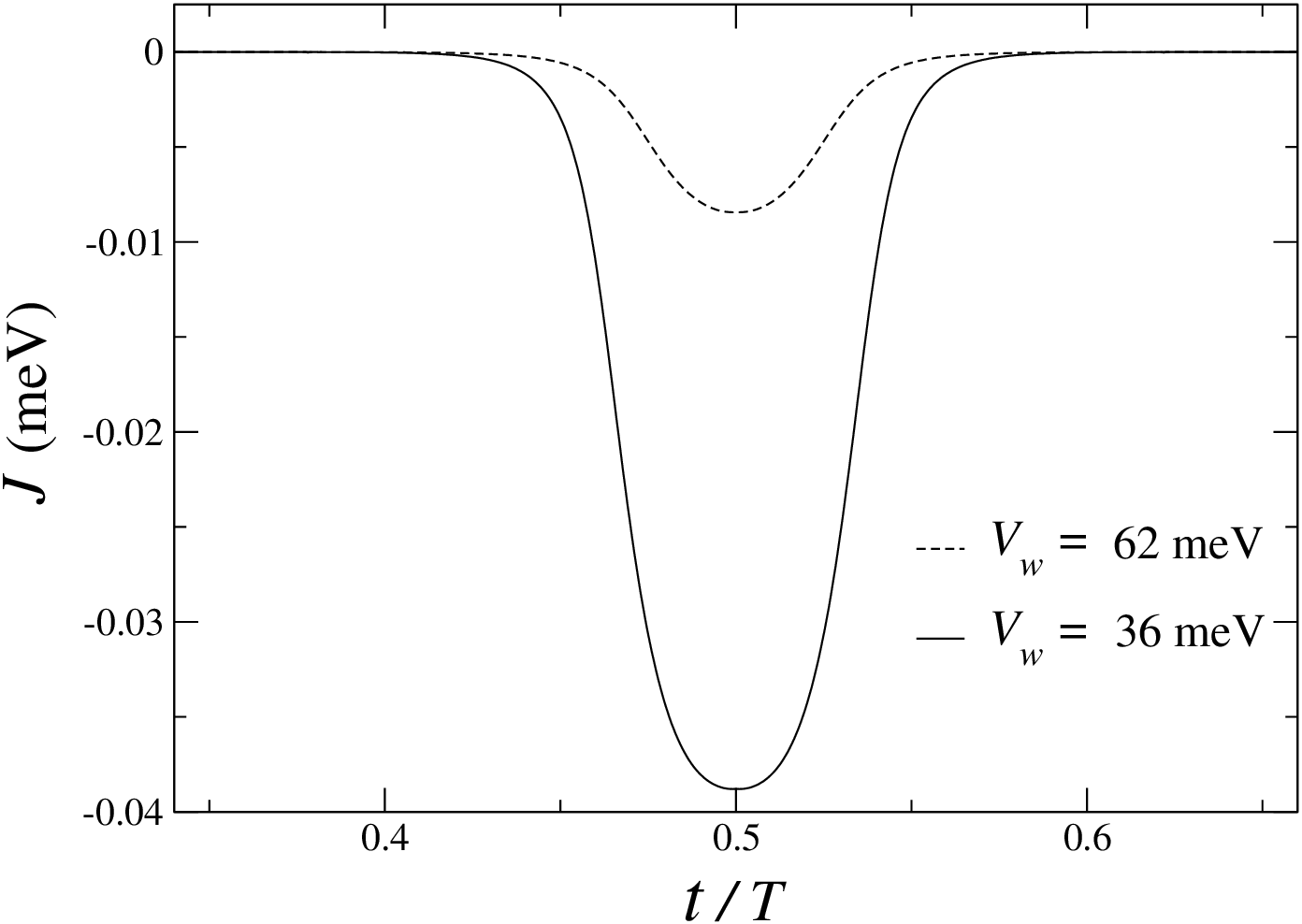}
\caption{Exchange energy as a function of time, for two different
well depths ($V_{w}$=36 meV, $V_{w}$=62 meV) and a SAW potential
amplitude of $V_{o}$=10 meV.}
\end{center}
\end{figure}

\section{Summary}

In summary, we have presented and investigated a scheme to produce
entangled states for two electrons utilizing a SAW. One electron
is carried by the time-dependent SAW potential along a
semiconductor quantum wire where a second electron is bound in a
quantum well. The Coulomb interaction induces entanglement between
the two electrons that can cover the full range from zero to full
entanglement, depending on SAW potential amplitude and the shape
of the confining potential. There are two regimes of interest,
depending on the SAW and well parameters.

The first is when there is a significant difference between
transmission probabilities for singlet and triplet states. In this
regime, entanglement generation may be interpreted as a
spin-filtering effect in which the singlet component of an
initially unentangled state has a higher transmission probability
than the triplet due to spin-dependent scattering. This gives
maximal entanglement ($C\sim 1$) for resonant singlet tunnelling
with full transmission for the singlet case and almost full
reflection for the triplet case. ``Fuzzy'' position measurement
(possibly through charge sensing\cite{delft1,delft2,delft3}), just resolving whether the
outgoing electron is transmitted or reflected, would be needed to
make this useful entanglement. The measurement result would
identify and herald the form of entangled state produced in each
run of an experiment.

The second regime occurs for the parameters chosen such that there
is approximately full transmission for both singlet and triplet
cases. Within this regime the transmitted electron in a SAW
minimum can be left in an excited state which in some cases is
different for singlet and triplet and in some cases the same (or
approximately so). In the latter cases, concurrence is given by a
simple expression involving the relative phase difference between
the transmitted SAW potential minimum wave functions arising from
singlet and triplet. This demonstrates maximal entanglement when
the phase difference is $\pi/2$ and we have identified a
physically reasonable set of parameters for which this occurs. For
other cases, the concurrence cannot reach the unitary limit and
can fluctuate significantly due to spin-dependent resonance
effects when the electrons interact. In this regime of near full
transmission, the concurrence is low when the transmitted SAW
minima wave functions are significantly different from each other
for singlet and triplet cases, becoming zero in the limiting cases
when these wave functions are orthogonal.

The physical system we have considered and the parameter ranges we
have investigated suggest that it should be experimentally
possible to produce useful entanglement between a travelling and a
trapped electron, using a SAW. This could be achieved either by
sensing whether the outgoing electron is transmitted or reflected,
or by working in a regime where there is essentially complete
transmission.

\section{Acknowledgments}

GG acknowledges funding from EPSRC and AR acknowledges support from the Slovenian Research Agency under contract PI-0044. This work was supported by the
UK Ministry of Defence.

\end{document}